\documentclass{emulateapj}

\usepackage{graphicx}
\usepackage{natbib}
\usepackage{color}

\newcommand{\kms}{km s$^{-1}$}

\newcommand{\mhz}{$\mu$Hz}
\newcommand{\teff}{$T_{\rm eff}$}

\newcommand{\si}{$\sim$}
\newcommand{\msol}{M$_{\odot}$}
\newcommand{\rsol}{R$_{\odot}$}

\newcommand{\age}{$\tau$}
\newcommand{\mh}{$[$M/H$]$}
\newcommand{\logg}{$\log g$}

\newcommand{\trat}{$T_{BA}$}

\newcommand{\ka}{$K_A$}
\newcommand{\kb}{$K_B$}

\newcommand{\per}{$\Pi$}

\newcommand{\ra}{$R_A$}

\newcommand{\masini}{$M_A \sin^3 i$}
\newcommand{\mbsini}{$M_B \sin^3 i$}

\shorttitle{$\delta$ Scuti stars in eclipsing spectroscopic binary systems}
\shortauthors{Creevey et al.}

\begin{document}


\title{Constraining the properties of delta Scuti stars using spectroscopic eclipsing binary systems}


\author{O.~L.~Creevey\altaffilmark{1,2}} 
\email{orlagh@iac.es}
\author{T.~S.~Metcalfe\altaffilmark{3}}
\author{T.~M.~Brown\altaffilmark{4}}
\author{S.~Jim\'enez-Reyes\altaffilmark{1}}
\author{J.~A.~Belmonte\altaffilmark{1,2}}


\altaffiltext{1}{Instituto de Astrof\'{i}sica de Canarias, C/ V\'ia L\'actea s/n,
E-38200 Tenerife, Spain.}
\altaffiltext{2}{Universidad de La Laguna, Avda. Astrof\'isico 
Francisco S\'anchez s/n, 38206 La Laguna, Tenerife, Spain.}
\altaffiltext{3}{High Altitude Observatory/National Center 
for Atmospheric Research, Boulder, Colorado 80301, USA}
\altaffiltext{4}{Las Cumbres Observatory Global Telescope Network Inc.,
6740 Cortona Dr. Suite 102, Santa Barbara, CA 93117, USA.}

\begin{abstract}

    Many stars exhibit stellar pulsations, favoring them for asteroseismic analyses.
    Interpreting the oscillations requires some knowledge of the 
    oscillation mode geometry (spherical degree, radial and azimuthal orders).
    The $\delta$ Scuti stars (1.5 -- 2.5 M$_{\odot}$) often show just one or few pulsation
    frequencies.
    Although this may promise a successful seismological analysis,
    we may not know enough about {either the mode} or the star to 
    use the {oscillation frequency} 
    to improve the determination of the stellar model, or probe the star's structure.
    For the observed frequencies to 
    be used successfully as seismic probes of these objects, we 
    need to concentrate on stars for which we can reduce the 
    number of free parameters in the problem, such as binary systems
    or open clusters.
    We investigate how much 
    our understanding of a $\delta$ Scuti star 
    is improved when it is in a detached eclipsing binary system
    instead of being a single field star.
    We use singular value decomposition to 
    explore the precision we expect in stellar 
    parameters (mass, age and chemical composition) for both cases.
    We examine how the parameter uncertainties propagate to the 
    luminosity -- effective temperature diagram and 
    determine when the 
    effort of obtaining a new measurement is justified. 
    We show that for the single star, a correct identification of the oscillation
    mode is necessary to produce strong constraints on the 
    stellar model properties, while for the binary system the observations without the pulsation 
    mode provide the same or better constraints on the stellar parameters.
    In the latter case, the strong constraints provided by the binary system not only allow us to 
  detect an incorrectly-identified oscillation mode, but we 
  can also constrain the oscillation mode geometry 
  by comparing the distribution of possible solutions with and 
  without including the oscillation frequency as a constraint.
\end{abstract}





\section{Introduction}

    $\delta$ Scuti stars are a class of pulsating objects located 
    on the Hertzsprung-Russell (H-R) diagram  
    on and near the main sequence (MS) where it intersects the 
    classical instability strip
    (see Breger 2000
    for a review).  
    They are 1.5 - 2.5 $M_{\odot}$ stars often pulsating in one
    dominant oscillation mode or in many lower-amplitude modes.
    Originally they were thought to be 
    very interesting targets for an asteroseismic analysis because 
    1) their pulsations are easily detected with ground-based
    telescopes, and 2) we 
    understand the structure of MS stars relatively
    well (e.g. Metcalfe et al. 2009). 
    Consequently, asteroseismology 
    of these stars was thought to have the potential to
    probe the details of the interior, such as energy
    transport mechanisms and convective core overshoot, as well as 
    less well-understood phenomena such as rapid rotation
    \citep{fea07,mac07}.
    
    Unfortunately, successful asteroseismic analyses are infrequent 
    for several reasons.
    1)   Observationally, the labeling of each measured frequency with 
    its geometric characteristics (mode identification) is not trivial.
    {For example, rapid rotation
    causes each of the mode
    degrees $\ell$ to split into ($2\ell+1$) components with
    different azimuthal order ($m$) \citep{gou05}.  
    Also some of the stars are sufficiently 
    evolved to show mixed modes \citep{met10}, thus complicating the analysis.}
    2) {Theoretically, it is difficult to find a unique model to match 
      the 
    set of observed frequencies, even with additional 
    observational constraints.  
    With the fundamental stellar properties poorly constrained, 
    use of the mode as a "seismic probe" is severely limited. 	}
    {For example, the multi-periodic $\delta$ Scuti star FG Vir, has 
      been the target of many 
     observational campaigns, allowing the detection of more than 75 frequencies
    \citep{bre05}, and the identification of 12 modes \citep{das05,zime06}.  
    Even with these observational constraints, the stellar parameters such as mass and helium
    mass fraction remain uncertain \citep{gb95,vis98,tem01,kir05}.
    A similar case is XX Pyx (see \citealt{han02} for a review).  
    \citet{han02} noted that even
    searching among more than 
    40,000 models in a three-dimensional parameter space, 
     \citet{pam98} could not find a model that matched the observations.}
    
    The lack of successful asteroseismic analyses has
    encouraged authors to reconsider their approach.
    On the one hand, much progress has been made 
    in observational mode identification,
    while on the other, authors are beginning to study these stars in systems where 
 	the number of free parameters is reduced \citep{lb00, ah04, mac06, cos07}.
   For example, open clusters provide constraints on the age and metallicity
    \citep{fox06,sae10}, while 
    multiple systems allow precise determinations of the mass and radius
    \citep{esc05,hek10,mac10}.
    The ideal pulsating star would be 
    part of an eclipsing binary within an open cluster 
    \citep{are07,cre09,ste10,tal10}.

    Using asteroseismology to probe the interior of a star
    requires the global properties to be known quite well.
    For example, the mass should be known to 1-2\%.
    Fortunately, the measurable quantities from a binary system can provide strong constraints
    on the properties of the component stars.
    If the binary is an eclipsing and double-lined spectroscopic system (SB2), the 
    absolute values of the masses and radii can be determined with 1-2\% 
    precision e.g. 
    \cite{rib99,las00}.

    {Many new detached eclipsing binaries are being discovered by several
    satellites dedicated to photometric monitoring of stars 
    (MOST \citealt{mat04},  CoRoT \citealt{bag02}, 
     {\it Kepler} \citealt{bor03}),
     as well as numerous ground-based surveys
     \citep{hen98,bc00,bak02,pol06}.
    Some of these systems have components that exhibit
    stellar oscillations \citep{mac10,hek10}.  }
    We will soon be faced with the problem of choosing which systems to study.
    
    In this paper, we use singular value decomposition (SVD) to study the
    information content of stellar systems.  
    This theoretical investigation quantifies the increase in astrophysical information realized by 
    studying pulsating stars in detached eclipsing SB2 binaries (binary) 
    instead of single pulsating field stars (single star),
    especially for those cases where mode identification is difficult.
    We assume that in a detached system, the stellar 
    structure is unaltered by the binarity.
    The particular questions we address are:
     \begin{itemize}

\item[-] What is the role of each observable for a single star,
and how do these roles change if the star is observed in a binary system?

\item[-] With the additional constraints for the binary 
system, how does the error box in the H-R
diagram compare to that of a single star?  
If the mode geometry is successfully identified through observational 
methods, how does this error box change?

\item[-] Given the typical observational errors, 
what are the expected uncertainties in the model
parameters for the single star and the binary system? 

\item[-] Are the constraints on the stellar parameters sufficient to 
distinguish between possible solutions when a mode is incorrectly identified?

\end{itemize}

In Section~\ref{sec:methods} we introduce the definitions, models, observations and method
necessary to understand the rest of this work. 
In Section~\ref{sec:singlestar} we study the roles that the observations 
play for determining the stellar model solution 
for both a single star and a binary stellar system, and 
we highlight which observations are most informative. 
In Section~\ref{sec:dscuti_sgldbl} we then discuss the parameter 
uncertainties, the error ellipses, and the error box in the H-R diagram, 
which show more specifically which observations are capable or incapable of reducing 
these uncertainties.  
Finally in Section~\ref{sec:five} we use simulations to show that an incorrectly-identified
oscillation mode can be detected in the binary system, but not in the single star.
We also show that the binary observations alone can 
constrain and even identify the mode geometry by studying the distribution
of model solutions without observational constraints
on the mode.

\section{Models, observations and methods}\label{sec:methods}
Our method follows that of \citet{bro94} who used 
SVD as a diagnostic tool to investigate how useful oscillation 
frequencies are as constraints on stellar parameters in a visual binary system, 
using $\alpha$ Cen A as an example.
This work was followed up in \cite{cre06} and \cite{cre07} and 
the techniques are also presented in \cite{pre92} and \cite{mm05}.
We refer the reader to these papers for details on the method.

\subsection{The distinction between parameters and observables}
{We define a {\it parameter} as an input characteristic to a stellar model, for example, mass and age.  These are the 
quantities that we wish to determine.
An {\it observable} is an output quantity from a stellar model given a set of parameters e.g. 
a radius, metallicity or a photometric magnitude $V$.
We compare the theoretical observables to the real observations in order
to retrieve the parameters.
For example, 
we can measure the effective temperature of a star (observation) and its error, and use
these to retrieve the mass and age (parameters) for a stellar evolution 
model by comparing the observed effective temperature  
to the theoretical/model effective temperature (observable).}

\subsection{Stellar parameters and models\label{sect:models}}

\begin{figure}
\center{\includegraphics[width = 0.48\textwidth]{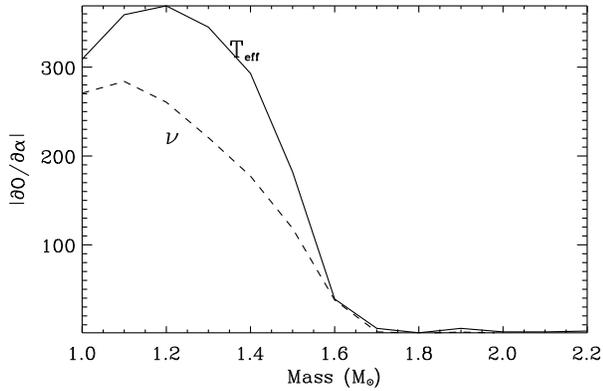}}
\caption{The dependence of the most sensitive observables 
(effective temperature and an oscillation mode) on the convective envelope mixing-length
parameter for a range of stellar masses.
\label{fig:alpharesp}}
\end{figure}

    We describe a single $\delta$ Scuti star by a set of parameters {\bf P}. 
    These are  
    mass $M$, age $\tau$, 
    rotation (we use rotational velocity $v$), initial 
    hydrogen $X$ (or helium $Y$) and heavy metal $Z$ mass fraction where $X+Y+Z = 1$, 
    and mixing-length parameter $\alpha$, 
 where applicable.
     Figure~\ref{fig:alpharesp} shows the derivatives of the two most sensitive observables 
     (the effective temperature and an oscillation mode) with 
     respective to $\alpha$ for a range of stellar masses.  
     For values above 1.7 M$_{\odot}$, we can clearly see that the considered 
     observables are insensitive to $\alpha$, and can be ignored.
    The Aarhus STellar Evolution Code (ASTEC) \citep{jcd08} is used to calculate the 
    stellar evolution models. 
    The ASTEC code uses the equation of state of \citet{egg73}, without Coulomb corrections, 
    and the OPAL opacities \citep{ir96}, supplemented by Kurucz opacities at low temperatures.
    The nuclear reaction rates come from \citet{bp92}, convection is described by the mixing-length
    theory of \citet{bv58}, convective core overshooting is included with 
    $\alpha_{\rm ov}$ set to 0.3, and
    diffusion effects are ignored.   
    This code uses the stellar parameters described above 
    as the input ingredients
    and returns
    a set of global stellar properties such as radius $R$ 
    and effective 
    temperature \teff\ (observables), 
    and the interior profiles of the stellar mass, 
    density and pressure.
    Oscillation frequencies are calculated
    using MagRot \citep{gt90, bt06}. 
    The distance $d$ is also included as a 
    stellar parameter, and coupling this 
    with $R$, $M$, \teff\ and the metallicity  \mh, allows us to calculate magnitudes
    using 
    model atmospheres \citep{lej97}. 

    For a binary system, the additional model parameters are the system properties:
    orbital semi-major axis $a$, orbital eccentricity $e$, 
    longitude of periastron $\omega$,
    systemic velocity $\gamma$, and orbital inclination $i$ (note that $i$ will always
    denote inclination unless otherwise specified).
    Both stellar components of a binary system
    share the parameters $\tau$, $X$ and $Z$,
    so the individual stars differ only in
    $M$ and $v$.
    We shall use the subscripts 'A' and 'B' to denote the components of the binary.
    The parameters of the model system are given in 
    Table \ref{tab:parameters}, and while
    these are not based on any particular binary system, there are known
    systems whose masses approximate these, such as
    HD~172189, HD~26591, HD~42083.
    We also note that the system stars pursue non-synchronised rotation,
    compatible with our earlier assumption that the stars evolve "individually".

    \begin{table}
      \centering
      \caption{\label{tab:parameters}System Parameters}
      \begin{tabular}{llllllllll}
	\hline\hline
	Parameter  & Value ($P_j$)&&Parameter  & Value ($P_j$)\\	\hline
	$M_A$ & 1.80 $M_{\odot}$ 	&&$v_A$ & 100.0 km s$^{-1}$ \\
	$M_B$ & 1.70 $M_{\odot}$ &&$v_B$ & 60.0 km s$^{-1}$\\
	$X$ & 0.700 &&$Z$ & 0.035\\
		$\tau$ & 0.700 Gyr &&$d$ & 200 pc \\
	$a$ & 0.150 AU &&$i$ & 85.6$^{\circ}$ \\
	$e$ & 0.0  &&	$\gamma$ & 0.0 km s$^{-1}$ \\
	($\omega$ & 0.0)\\
	\hline
      \end{tabular}
    \end{table}

\subsection{System observables and observations}
    The observables are the measurable quantities of the system.
    These include things such as $R$ (from interferometry for example),
    \teff, 
    \mh,
    gravity \logg, and parallax
    $\pi$ for a single star.
    For a binary system, additional observables include
    effective temperature ratio \trat\ (= $T_B/T_A$), 
    relative radii $R_{BA}$, 
    semi-amplitudes of radial velocity curves
    \ka\ and \kb, and orbital period \per. 
    We note that  the radial velocity or light curve measurements are also considered observables,
    however, in this work  the more intuitive derived values (\trat\, \masini, \mbsini, $\omega$ etc.) are used.

Using model atmospheres \citep{lej97} with $d$ and $R$, we calculate the 
photometric magnitudes ($r$, $i$, and $z$) using the 
SDSS filter system \citep{yor00}.  These atmospheres take reddening into
account.  
We subsequently calculate the colors from these magnitudes, 
although we note that distance $d$ is not necessary to obtain 
approximate color indices.
The colors we include in our analysis are $(r-i)$ and $(i-z)$ and the magnitude is $r$.
{For the binary, the photometric observables are calculated from 
blended spectra.  
Two or more blended colors yield estimates of the individual and relative
quantities of the component properties.  }
{The colors of the individual components can be disentangled \citep{koc60,sem01,cre05}.
However, obtaining component colors with the same precision as the system can
only be achieved with high quality multi-color  light curves.}

From the light curve of an eclipsing binary system we obtain
$R_{BA}$, $T_{BA}$, $i$ and $\Pi$.  
We use these derived values from the modeling of the binary system 
for simplicity and to understand
the errors in terms of what is quoted in the literature.
The derived observables are valid, as long as we 
correctly propagate the  uncertainties from the light curve data.

Spectroscopy provides the observables $v_A \sin i$ and $v_B \sin i$ 
(projected rotational velocity)
and the atmospheric parameters $\log g$, $T_{\rm eff}$ and $[$M/H$]$.
{An observed spectrum for a binary system is 
the sum of the spectra from the two stellar components.   
Many observational techniques exist to disentangle spectra e.g. \citet{had95,had09},
and these techniques have been successful, see \cite{cre09} for a specific
case using various techniques to determine the individual component \teff\ for HD~172189.
We assume that we can successfully disentangle the spectra and obtain 
effective temperatures for both stars while 
using the photometric \trat\ as a constraint.}

A time series of spectroscopic measurements yields 
a time series of radial velocities, and 
modeling these data yields
$M_A \sin^3 i$, $M_B \sin^3 i$, $a \sin i$, 
$e$, $\omega$ and $\gamma$. 
Coupling these values with $i$ and $R_{BA}$ (from photometry) provides
the absolute values of the radius.

A time series of radial velocity or photometric measurements provides
the frequency(ies) $\nu$ of the modes that are excited in the star.
However, the frequencies are only useful if we know the 
oscillation mode geometry
(the degree $\ell$, the azimuthal order $m$ and the radial order $n$
of the wave).
This so-called "mode-identification" can be achieved by either 
studying the time series of 
spectroscopic line-profiles of the star \citep{ken90,hor96,bal00,ba03,zim06},
by using multi-color photometry and comparing the oscillation
amplitudes and the phases of the frequencies in various filter bands 
\citep{gar00,dup03,das07},
or by using the screening effects of the eclipses in eclipsing binaries \citep{bir05,gam05}.
Many $\delta$ Scuti 
pulsators have only one dominant oscillation frequency
and so the assumption is that we measure at least one precise frequency.

\begin{table}[t]
\begin{center}
\caption{Observables and Errors \label{tab:obsers}}
\begin{tabular}{lccccccccrrrrrrrr}
\hline
\hline
&&Observable ($B_i$) & Error ($\epsilon_i$)\\
&& &Figs.~\ref{fig:signif_sets_s},\ref{fig:signif_obs_s},\ref{fig:signif_sets2} && 
Figs.~\ref{fig:dscuti_svd_sgla},\ref{fig:ellipses} & 
Figs.~\ref{fig:binsglpar}-\ref{fig:fitage}\\
\hline
$R_A$ (\rsol) &&1.95&  0.02 &\hspace{.5cm}& 0.01 & 0.04\\
$R_B$ (\rsol) && 1.81& 0.04 & & 0.01 & 0.04\\
$T_{BA}$ &&  0.97&0.05 && 0.05 & 0.05\\
$i$ ($^{\circ}$) &&85.6& 1.0 && 1.0 & 0.3\\
$T_{\rm eff A}$ (K) && 6965& 100 && 100 & 100 \\
$v_A \sin i$ (\kms) && 99.7& 2.5 && 2.5 & 2.5\\
$v_B \sin i$ (\kms) &&  59.8&2.5 && 2.5 & 2.5\\
$\Pi$ (yrs) && 3.1053e-6 & 1e-6 && 1e-6 & 1e-6\\
$M_A \sin^3 i$ ($M_{\odot}$) && 1.78 & 0.06 && 0.10 & 0.04\\
$M_B \sin^3 i$ ($M_{\odot}$) && 1.69 &0.05 && 0.10 & 0.03\\
$[$M/H$]$ (dex) &&  0.31 & 0.05 && 0.05 & 0.05\\
$\log g$ (dex) &&  4.1 & 0.5 && 0.5 & 0.5\\
$\pi$ (mas) && 5.0& 0.5 && 0.5 & 0.5\\ 
$r$ (mag) &&  8.16 & 0.02 && 0.02$^{\dagger}$ & 0.01\\
$(r-i)$ (mag) &&  0.178 &0.003 && 0.005$^{\dagger}$ & 0.001\\
$(i-z)$ (mag) && 0.009&0.003&& 0.005$^{\dagger}$ & 0.001\\
$\nu$ (\mhz) && 1108.9 & 1.3 & & 1.3$^{\dagger\dagger}$ & 1.3\\
\hline
\hline
\end{tabular}
\end{center}
$^{\dagger}$ 0.5 mag for Fig. \ref{fig:dscuti_svd_sgla} top panel.
$^{\dagger\dagger}$ 1e4 \mhz\ for the black ellipse in Fig.~\ref{fig:ellipses} 
\end{table}

The values of the errors (c.f.~Table~\ref{tab:obsers}) 
are chosen by considering typical observations for such
systems, but also to study some limiting cases, {for example when the errors reach 
the limits of current techniques, or if the stars are bright (closer in distance).} 
We assume that the light curves cover a few orbital periods.  
{Spectroscopic data, however, are more time-consuming and the demand for observation time is 
high, especially for larger telescopes.}  
Thus, one may hope that spectroscopy during one full orbital period can be obtained, with some data points during 
subsequent orbits.  
Taking into account that one of the stars is pulsating, this will impact 
the analysis of both the photometric and spectroscopic data.
In particular the depths of the eclipses will need to be observed several times to remove the 
effects of 
the modulations due to pulsation, and this influences the derivation of 
\trat, R$_{BA}$, and $\Pi$, for example.
Spectroscopy will show clear line-profile variations, and deviations in the 
expected radial velocities due to pulsation.
Both of these will influence the determination of the orbital radial velocities to some 
extent, and these will in turn affect the errors in $M$, $R$ and $\Pi$. 
Table \ref{tab:obsers}, first column, lists the values of the observables calculated from
the stellar model defined by 
the parameters in Table~\ref{tab:parameters}. 
The subsequent columns show the corresponding errors used for this analysis.
We have considered errors ($\epsilon$) in mass-related and radius observables of the order of 1-3\%. 
The 3\% error should account for the effects of pulsation on the data. 
Errors of 100 K, 0.05 dex and 0.3 -- 1.0$^{\circ}$ for \teff, \trat, and 
$i$ are values typically found in the recent literature 
\citep{rib02, sou04, hil05, sou06, bru10}.
We use 2.5 \kms\ as the error on the projected rotational velocity 
(e.g. \citealt{cre09}).
Reducing the precision on these values influences the determination of
the rotational velocity of
the star.  
Unfortunately including rapid rotation is a complex matter for stellar models, and 
most stellar codes (including those used in this work)
consider only solid-body rotation which has minimal effect on the non-seismic observables.

\begin{table}
\begin{center}
\caption{Observable (OS) and parameter (P) sets used in this analysis.  
AS denotes additional observables
and the subscripts 'S' and 'EB' refer to single star and eclipsing binary system.  \label{tab:os}}
\begin{tabular}{ll}
\hline
\hline
{\bf OS}$_S$ =& $\{R, T_{\rm eff}, \log g, v \sin i, \mathrm{[M/H]}, \pi \}$ \\
{\bf OS}$_{EB}$ =& $\{R_{A},R_{B}, T_{BA}, i, T_{\rm eff\it A}, v_A \sin i, 
    v_B \sin i$, \\
& $\Pi, M_A \sin^3 i, M_B \sin^3 i, \mathrm{[M/H]}, \pi\}$\\
{\bf AS1} =& $\{\nu \}$\nonumber\\
{\bf AS2} =& $\{r, r-i, i-z \}$\nonumber\\
{\bf AS3} =& $\{r, r-i, i-z, \nu \} $\\
\\
\\
{\bf P}$_S$ = & \{$M,\tau,X,Z,v,i,d$\} \\
{\bf P}$_{EB}$ = & \{$M_A,M_B,\tau,X,Z,v_A,v_B,i,d,[a,e,\omega,\gamma]$\} \\
\hline\hline
\end{tabular}
\end{center}
\end{table}

The precision in the photometric values is typical for well-determined 
literature values.
We chose some limiting values ($\epsilon < 0.003$) because at these 
precisions
the colors begin to have an important impact for the determination of the 
parameters.
The frequency error of 1.3 \mhz\ corresponds to the frequency resolution of data from 
a one-week observational campaign.
We increase the values of these errors for some calculations (note in Table~\ref{tab:obsers}) to eliminate
their influence. 

An analysis of the influence that each observation has
(see next section) 
for different sets and errors,
and an investigation of the parameter uncertainties led us to 
choose four competitive sets of observables to study, and these
are given in the top part of 
Table~\ref{tab:os}.  {\bf OS} is the base
{\it observable set} used, and {\bf AS1}, {\bf AS2} and 
{\bf AS3} are the {\it additional sets}.
The subscripts 'S' and 'EB' denote 'single star' and 'eclipsing binary'
respectively.
For the remainder of this work we discuss our results in 
terms of each of these sets and we remove the bold text: 
OS, OS+AS1, OS+AS2 and OS+AS3.
We include the subscripts 'A' and 'B' on the observables and
parameters to denote
each component in the binary system when necessary, 
and the magnitudes and colors
also have a subscript 'EB' to emphasize that these are 
binary system (blended) observables.

\begin{figure}
\center{\includegraphics[width = 0.48\textwidth]{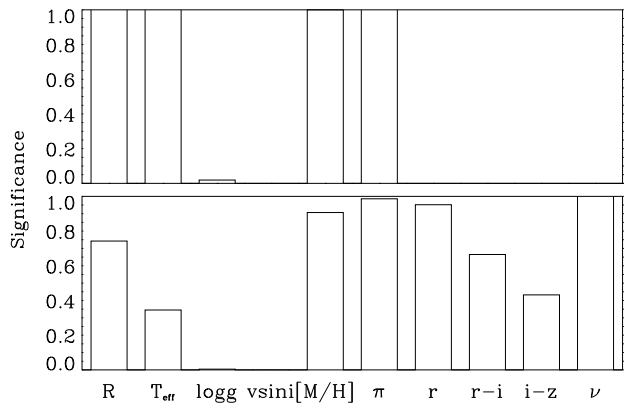}}
\caption{Significance of observables sets without colors, magnitudes and the
oscillation mode (OS$_S$ --- top) and including these data (OS$_S$+AS3 --- bottom) for
a single star.}
\label{fig:signif_sets_s}
\end{figure}

We do not include  $a \sin i$ as an observable because it does note provide independent 
information.  We can calculate it
from three of the following:
$M_A \sin^3 i$, $M_B \sin^3 i$, $q \equiv K_A/K_B = M_B/M_A$, and $\Pi$ using
Kepler's Third Law.

\subsection{Singular value decomposition method\label{sect:svd}}

    SVD is the decomposition of any $M\times N$  matrix {\bf D} into 
    3 components: ${\bf U}$, ${\bf V}$ 
    and ${\bf W}$ given by ${\bf D} = {\bf U W V^T}$.
    ${\bf V^T}$ is the transpose of ${\bf V}$ which is an 
    $N \times N$ orthogonal matrix that contains the
    {\it input} basis vectors for ${\bf D}$ or the vectors
    associated with the parameter space.
    ${\bf U}$ is an $M \times N$ orthogonal matrix that contains the
    {\it output} basis vectors for ${\bf D}$, or the vectors
    associated with the observable space.
    ${\bf W}$ is a diagonal matrix that contains the
    {\it singular values} of ${\bf D}$.  

    The key element to this work is the description of the matrix {\bf D}, which
    we call the design matrix and each element is a 
    partial derivative of each
    of the observable with respect to each of the parameters of the system,
    taking into account the measurement errors for each
    of the observables:
    \begin{equation}
      D_{kj} = \frac{\partial B_k}{\partial P_j} / \epsilon_k.
      \label{eqn:designmatrix}
      \end{equation}
    Here $B_k$ are each of the $k = 1, 2, ...M$ 
    observables of the system with expected 
    errors $\epsilon_k$,
    and $P_j$ are each of the $j = 1,2,...,N$ free parameters of the
    system (see section \ref{sect:models} 
    for discussion on the observables and the 
    parameters).
    
    By writing the design matrix with the measurement errors taken into account,
    we provide a quantitative description of the information content 
    of each of the observables for determining the stellar parameters
    and their uncertainties.

    Starting from an initial close guess of the solution ${\bf P}_0$, 
    SVD can be used as an inversion technique 
    to obtain the 
    true solution ${\bf P}_{\rm R}$ 
of the system.
    This is done by calculating a set 
    of parameter corrections ${\bf \delta P}$ 
    that minimizes some goodness-of-fit function:
    ${\bf \delta P = V\bar{W}^{-1}U^T \delta B}$, 
    where ${\bf \delta B}$ are the differences
    between the set of actual observations {\bf O} 
    and the calculated observables
    ${\bf B}_{\rm 0}$ given the initial parameters ${\bf P}_{\rm 0}$.
    ${\bf \bar{W}}$ is a modification of the matrix {\bf W} such that 
    the inverses of all values below a certain threshold are set to 0.
    The formal errors comprise the sum of all of 
    the ${\bf V}_q/w_q$, where each
    ${\bf V}_q/w_q$ describes the direction and magnitude to change 
    a parameter so that the true solution ${\bf P}_{\rm R}$ and formal 
    uncertainties can be given by
    \begin{equation}
      {\bf P}_{\rm R} = {\bf P}_{\rm 0} + 
      {\bf V\bar{W}^{-1}U^T \delta B} \left ( \pm \frac{{\bf V}_1}{w_1} 
      \pm  \frac{{\bf V}_2}{w_2} \pm ...
      \pm \frac{{\bf V}_N}{w_N} \right).
      \label{eqn:dscuti_er}
    \end{equation}
    The {\it covariance matrix} {\bf C} consequently assumes a  
    neat and compact
    form:
    \begin{equation}
      C_{jl} = \sum_{q=1}^N \frac{V_{jq}V_{lq}}{w_{q}^2},
      \label{eqn:covariance}
    \end{equation}
    and the square 
    roots of the  diagonal elements of the covariance matrix are the 
    theoretical parameter uncertainties.  Note that $\epsilon$ is reserved
    for {\it observational error} and $\sigma$ for {\it parameter uncertainty}:
    \begin{equation}
      \sigma_j^2 =  \sum_{q=1}^N \left ( \frac{V_{jq}}{w_q} \right )^{2}.
      \label{eqn:uncertainties}
    \end{equation}

Another useful property of SVD is the {\it significance} $S$ of an observable.
This can be quantified as a measure of the extent that a 1-$\epsilon$ change in
$B_k$ shifts the inferred parameters towards the 1-$\sigma$ error 
ellipsoid in parameter space.  
In this way, $S$ quantifies the impact that an observable has for the 
determination of the parameter solution:
\begin{equation}
S_k = \left ( \sum_{q=1}^N U_{kq}^2 \right )^{1/2}.
\label{eqn:signif}
\end{equation}
Because of the orthonormality of the decomposition matrices, the value of 
$S$ varies between 0 and 1.  A low $S_k$ implies that the observable
$B_k$ has relatively less influence on the solution, and that a change in the 
measurement will have little or no impact.
A high value of $S_k$ implies that this observable is important for the solution
and any change in the measurement will force a corresponding 
change in the solution.

Finally, the matrices {\bf U} and {\bf V} provide information about
the role
each of the observables plays in determining the stellar parameters. 
Each column vector of {\bf U} ($U_j$) is related uniquely to each column vector of
{\bf V} ($V_j$) and its importance in the solution is given by the corresponding 
singular value $w_j$ (cf. Fig \ref{fig:dscuti_svd_sgla}).  In section~\ref{ssec:dscuti_dermat} we elaborate on this
discussion.

\section{Binary system versus single star constraints}\label{sec:singlestar}

In this section we investigate the
roles that each 
observable plays for determining each parameter.
We study how these roles change when we include/exclude
certain observables and 
when we consider different values of the observational errors.

\subsection{Single star observables}

\subsubsection{Significance}

\begin{figure}[!h]
\center{
\includegraphics[width = 0.48\textwidth]{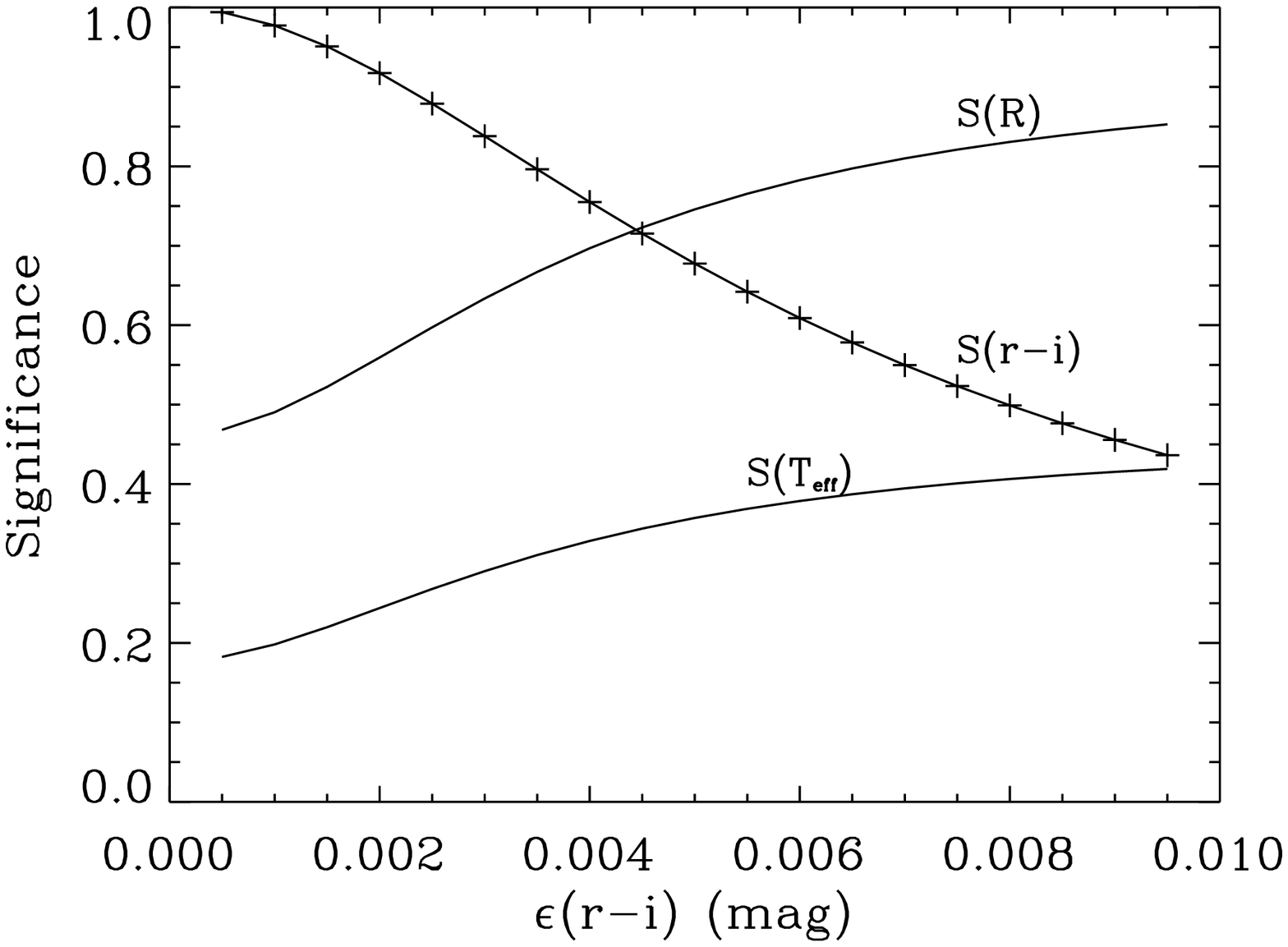}
\includegraphics[width = 0.48\textwidth]{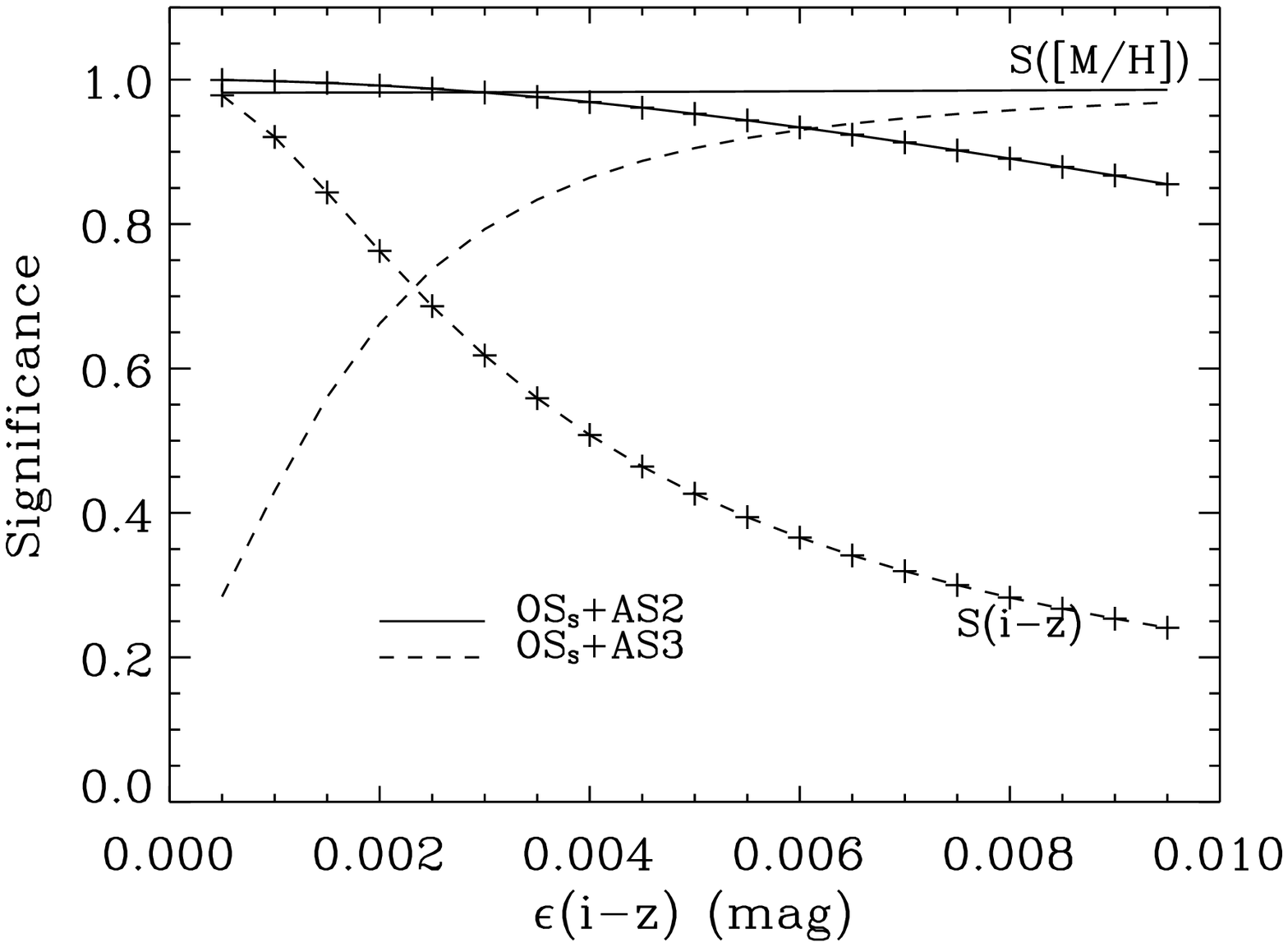}}
\caption{Change in observable significance $S$ as the precision in 
photometric information
changes for a single star.
The photometric observables are denoted by crosses on their lines.
{\sl Left:} 
Similar results are found for sets OS$_S$+AS2 and OS$_S$+AS3, as 
the precision is decreased in $\epsilon_{r-i}$.
{\sl Right:} 
Continuous and dashed lines
denote OS$_S$+AS2 and OS$_S$+AS3, respectively,
as the precision in $\epsilon_{i-z}$ is decreased.
 \label{fig:signif_obs_s}}
\end{figure}

Figure \ref{fig:signif_sets_s}
shows the significance (Eq.~\ref{eqn:signif}) of each of the observables
for a single star for OS$_S$ (top panel) and OS$_S$+AS3 (lower panel), using
the measurement errors from the second column in Table \ref{tab:obsers}.
Because there is no information about $i$ (inclination), the figure illustrates that
$v \sin i$ is not an effective constraint ($S=0$).
{Although, the value of $v \sin i$ imposes 
a lower limit on $v$, we can not determine its uncertainty.}
 \logg\ is also a weak constraint,  because 
both $R$ and $T_{\rm eff}$ provide similar information, but of better quality.
If either of these were not available, then $\log g$ would have a higher
significance.
$S=1$ implies that $R$, \teff, [M/H], and $\pi$ are all necessary observations
to constrain {\bf P$_S$}.
By including  the photometric information and the mode (lower panel), the most
noticeable change is the reduction in 
$S($\teff$)$, whose information is superseded mainly by $r$ and 
$(r-i)$. 
$S(\nu) = 1$ indicates that the mode has a strong influence on the determination of the solution.

Figure \ref{fig:signif_obs_s} shows how $S$ changes for  each observable
 as some measurement errors are varied.
The left panel shows $S(r-i)$, $S(R)$ and $S(T_{\rm eff})$ as 
$\epsilon_{r-i}$ is varied for OS$_S$+AS2.
Similar results are found for OS$_S$+AS3. 
$S(r-i)$ decreases as its error increases to values of about 0.004 mag, 
and, at this value, $S(R)$ begins to supersede $S(r-i)$.
$S(T_{\rm eff})$ also increases as the color measurement is more poorly
determined, although it never increases to more than 0.45.
The largest increase (or change) is seen only  
as $\epsilon_{r-i}$ reaches mmag precisions.

The right panel of Figure \ref{fig:signif_obs_s} shows the change in  
$S([$M/H$])$ and $S(i-z)$ as $\epsilon_{i-z}$ increases in value. 
In this case, no other observable is affected.
$(i-z)$  has a strong influence for
determining the parameter solution only at 
precisions of $\leq$ 0.002 mag; 
for $(r-i)$ the critical value is $\sim$0.004 mag (left panel).
The difference between the continuous and the dashed lines is the inclusion
of seismic data (AS3, dashed).
Without seismic data, both $[$M/H$]$ and $(i-z)$ are important observables, 
while including seismic data, 
$[$M/H$]$ has little impact when $(i-z)$ is important, 
and vice versa.
It is primarily the parameter $Z$ that these 
observables are responsible for determining.

\begin{figure*}[h!]
\center{\includegraphics[width = 0.99\textwidth,height = .3\textheight]{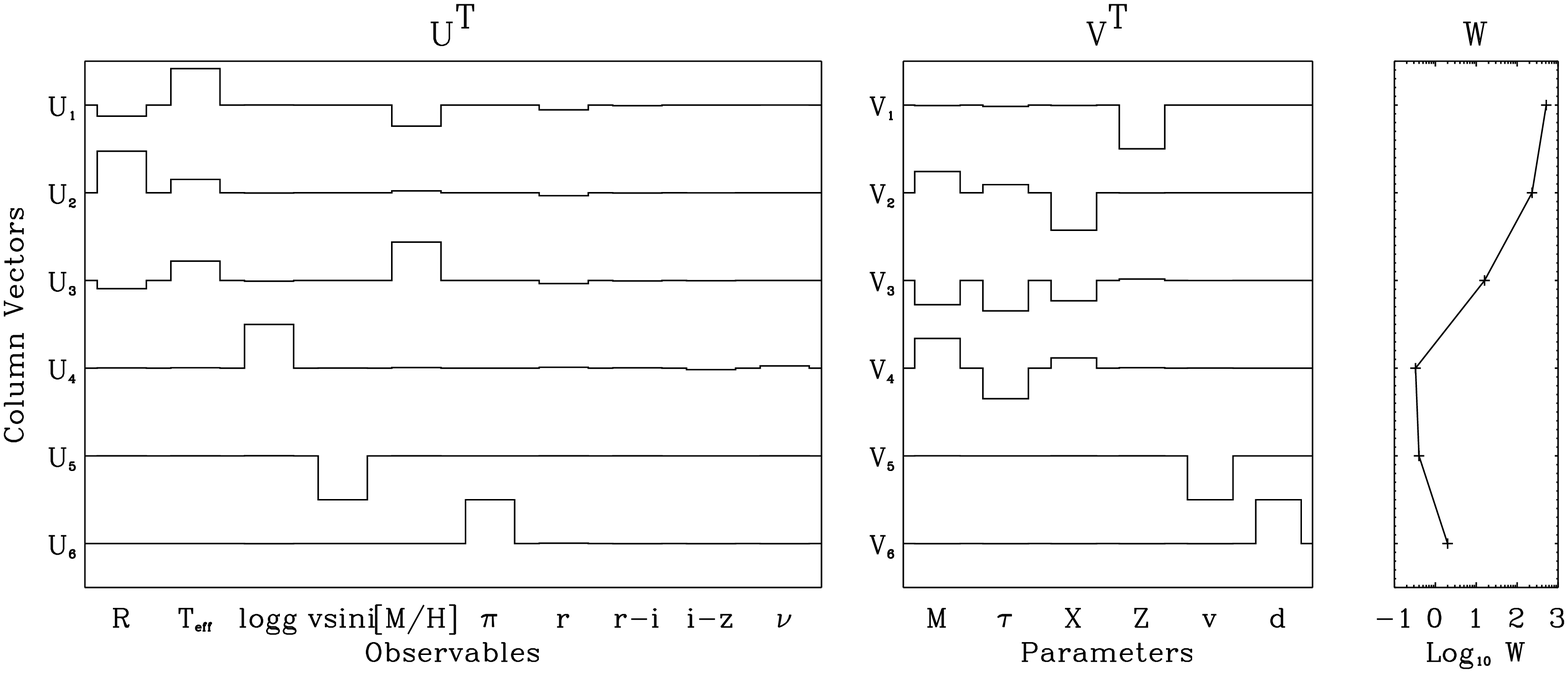}}
\center{\includegraphics[width = 0.99\textwidth,height = 0.3\textheight]{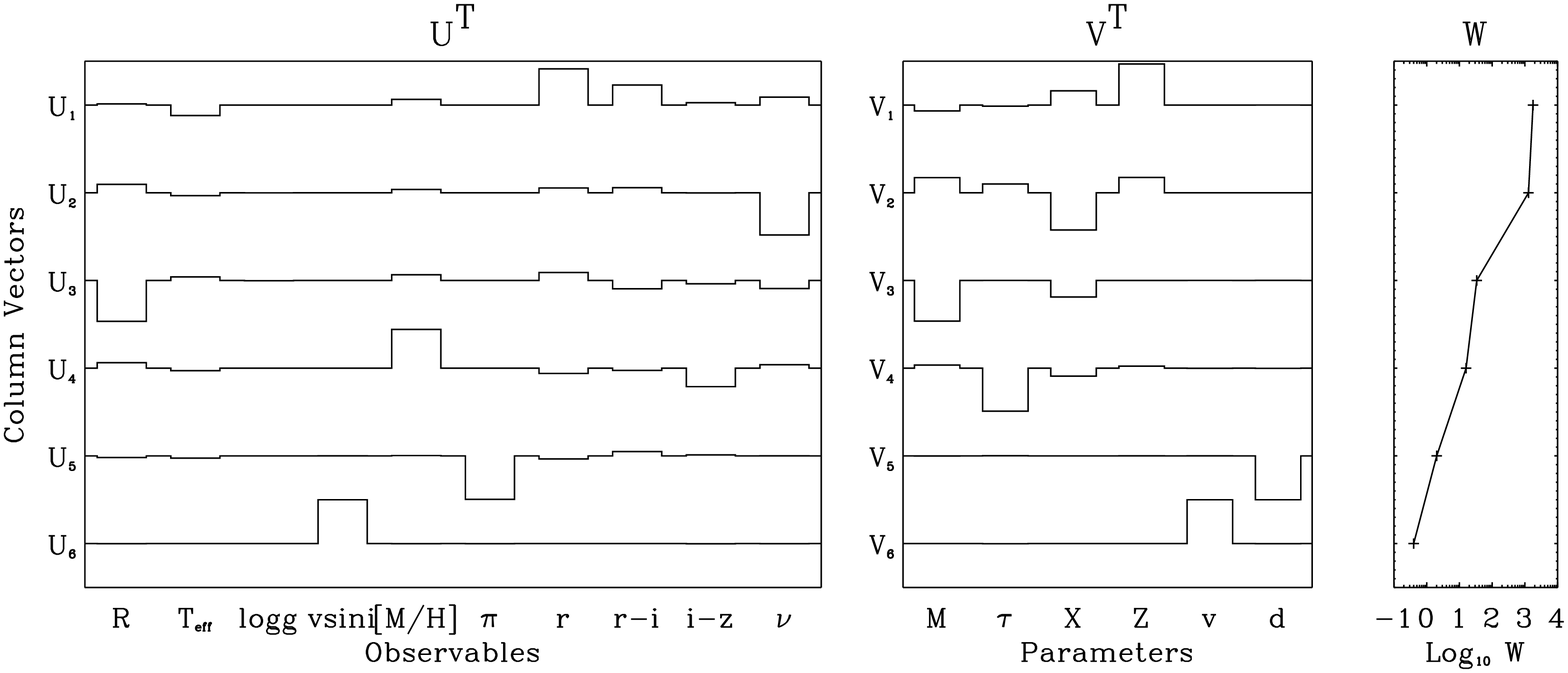}}
\caption{Decomposition matrices of a single star model using the errors given
in Table~\ref{tab:obsers}, third column.  The left, center and right panels are the
matrices U$^{\rm T}$, V$^{\rm T}$ and W respectively.
The difference between the panels is the 
addition of well-constrained photometric and seismic data in the lower panels.
These values are set to 0.5 mag and 1e4 $\mu$Hz respectively in
the upper panel.
\label{fig:dscuti_svd_sgla}}
\end{figure*}

\subsubsection{Transformation matrices \label{ssec:dscuti_dermat}}

The decomposition matrices of SVD show which observables determine each of 
the parameters of the system.  
These are sets of linear vectors where each vector
in {\bf U} is related uniquely to each vector in {\bf V}, 
and the relative importance of each vector
is given by the corresponding singular value $w_j$ in {\bf W}.  
In Figure \ref{fig:dscuti_svd_sgla}, 
we show the transpose of 
these decomposition matrices for the single star system, with the singular values represented in the
right-most figure of both panels.
These figures take into account the 
errors given in the third column of Table \ref{tab:obsers}, except 
for the upper panel
where the values for the colors
have been increased to 0.5 mag and the frequency error to 1e4 \mhz\
to eliminate their influences.

In Figure \ref{fig:dscuti_svd_sgla} top panel, $R$, \teff, and 
[M/H]
appear in the top three observable vectors indicating their importance for 
determining the combination of parameters, shown in vectors 
$V_1$, $V_2$ and $V_3$ (mass, age, chemical composition).
The next important observable (in $U_6$ --- here $w_6 > w_4$) is $\pi$, 
and this is uniquely 
responsible for determining the distance (in $V_6$).
Likewise, only the observable $v \sin i$ determines $v$, but poorly, 
and finally the observable
with least information is \logg\ ($U_3$ with smallest $w_3$), 
which contributes weakly to determining mass, age and chemical
composition.

In the bottom panel of Figure \ref{fig:dscuti_svd_sgla}, we have improved the errors in the photometric 
observables and the oscillation mode (third column in Table~\ref{tab:obsers}).  
These new observables
now play a dominant role for determining the parameters, as can
be seen from their positions in the top vectors, as well as an increase by 
a factor of 10 in the top singular values.
 Mass, age and chemical composition are much 
better constrained, mostly appearing in the top two panels.
The contributions from \teff\ and \logg\ have diminished considerably, and
the distance is now no longer determined uniquely by $\pi$.  It also depends weakly
on the photometric colors and magnitude ($V_5$ and $U_5$).
As in the top panel, $v \sin i$ is responsible for determining $v$ ($U_6$ and $V_6$).
By omitting the oscillation frequency from the set of observables, 
$R$
becomes the observable mostly responsible for determining $M$.

\subsection{Binary system observables: Significance \label{sec:dsstar}}

\begin{figure*}
\center{\includegraphics[width = 0.8\textwidth]{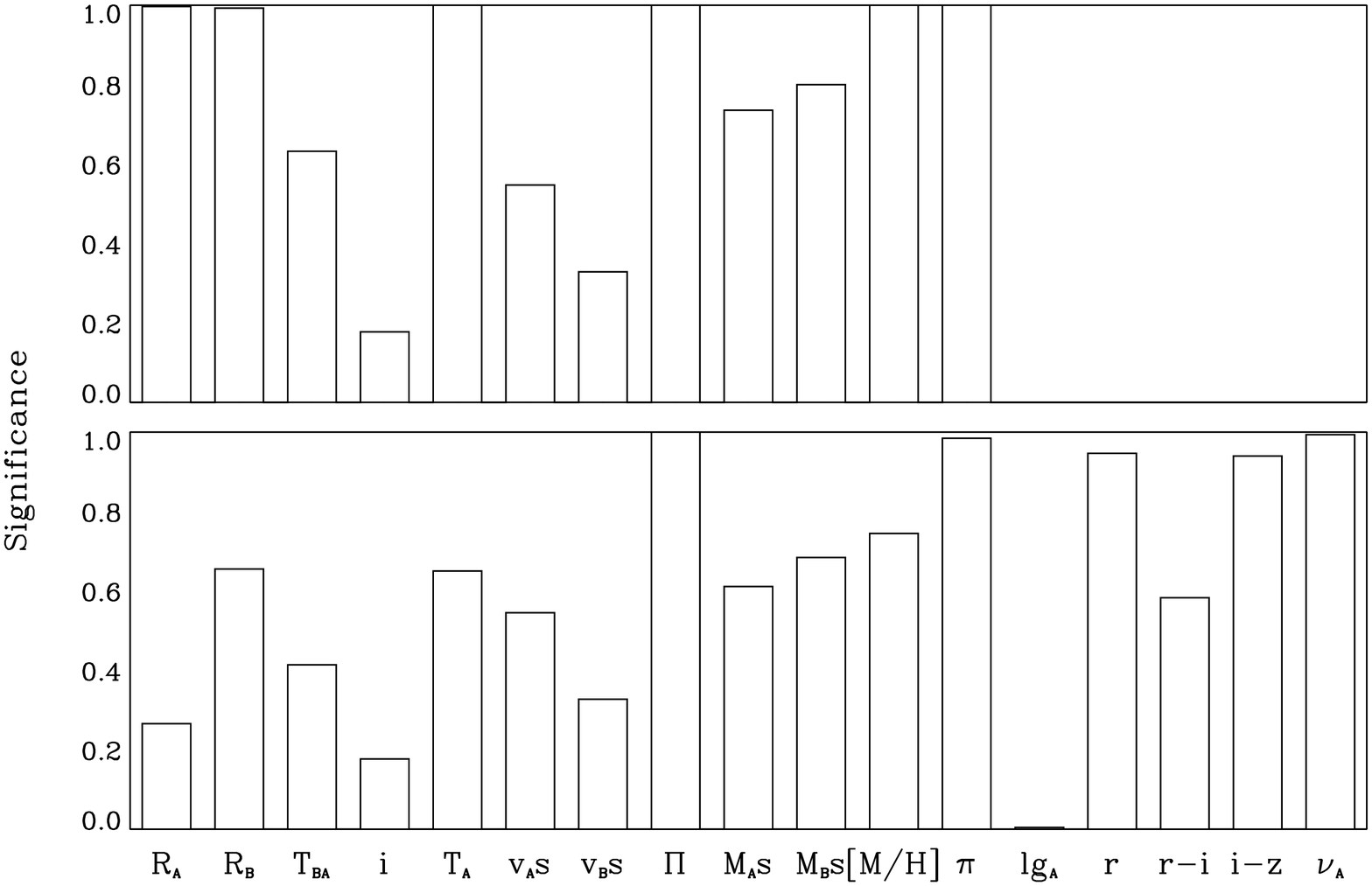}}
\caption{Significance of observables sets without/with photometric and seismic
data (OS$_{EB}$ /OS$_{EB}$+AS3) in top/bottom panels for
a $\delta$ Scuti star in a detached eclipsing binary system.
Note that we use an abbreviated labeling for the observables due to space constraints.
The errors are given in the second column of Table \ref{tab:obsers}.}
\label{fig:signif_sets2}
\end{figure*}

The top panel of Figure \ref{fig:signif_sets2} illustrates  {\bf S} 
for OS$_{EB}$ (binary system) using  the errors from the second {column} of  
Table~\ref{tab:obsers}. 
This figures shows that the most important quantities are
$R_A$, $R_B$, 
$T_{\rm eff\it A}$, $\Pi$, $[$M/H$]$,
$M_A \sin^3 i$, $M_B \sin^3 i$ and $\pi$. 
The lower panel shows {\bf S} for OS$_{EB}$+AS3.
Note how each of the $S$ values changes: including the magnitude, colors and oscillation
frequency reduces the importance of \ra, \teff$_{A}$, [M/H] and \trat, while ever so slightly
reducing $S(\pi)$.
{We note that $i$ appears deficient in information content.
An accurate and precise determination of $i$ is necessary to correctly scale the observed
values of \masini\ and \mbsini.  
In this case $\epsilon(i) = 1.0^{\circ}$, however, increasing its precision to $<$ 0.3$^{\circ}$ causes $S(i)$ to 
increase rapidly.}


\begin{figure*}
\includegraphics[width = 0.46\textwidth]{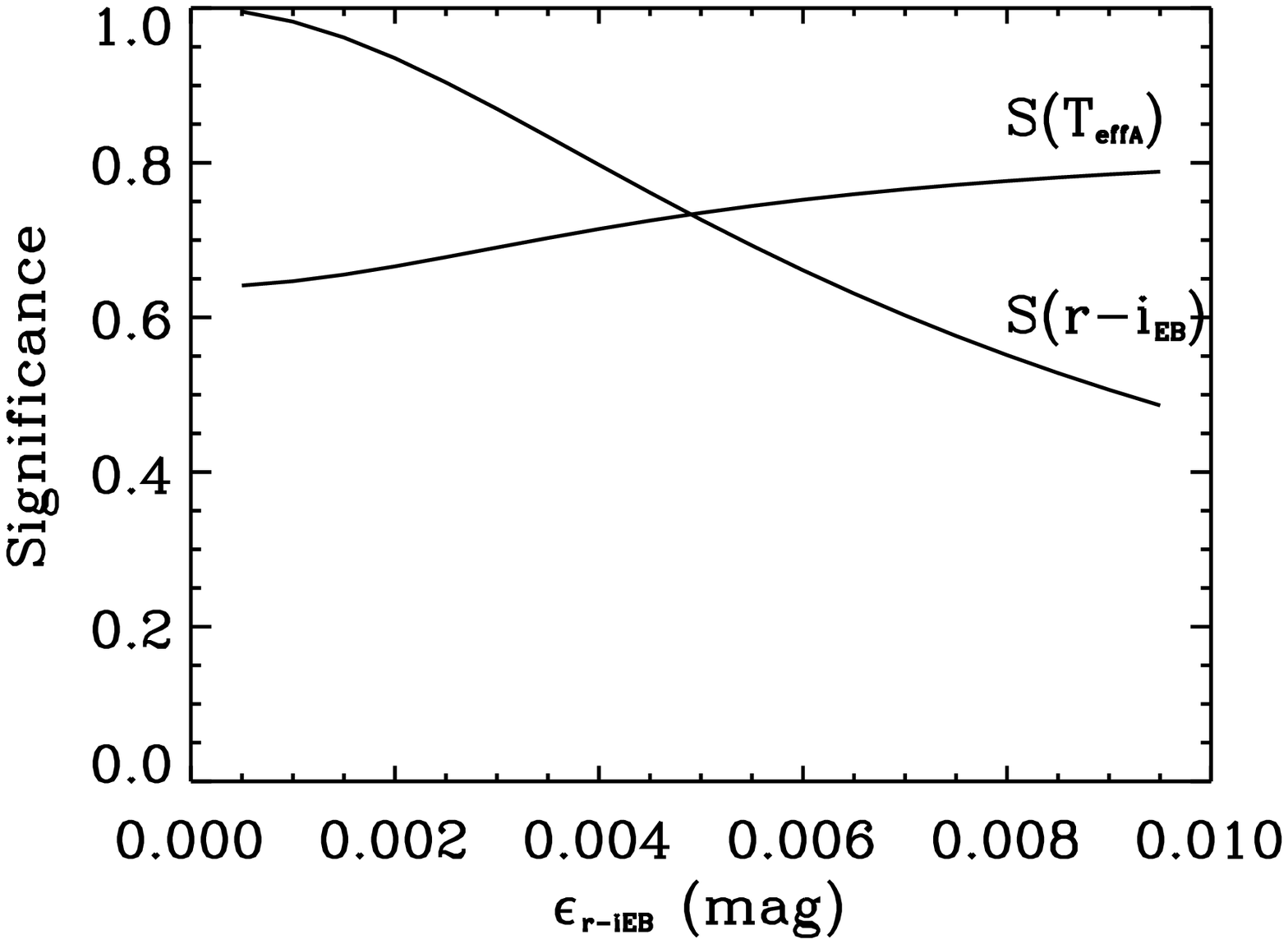}
\includegraphics[width = 0.46\textwidth]{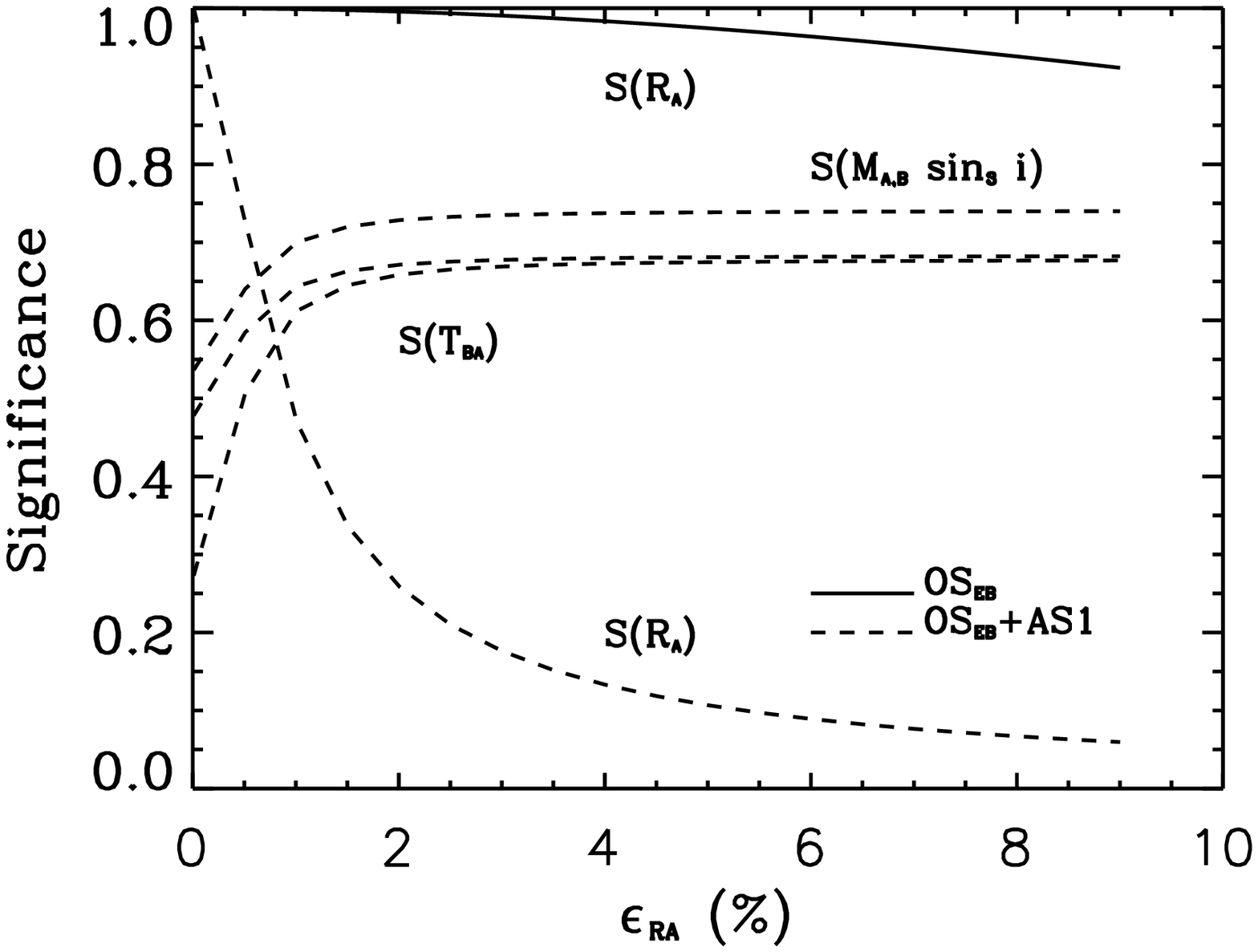}
\caption{The relative significance of some binary observables as the precision in their errors
changes.
{\sl Left:} $S({r-i})_{EB}$ and $S({T_{\rm effA})}$ change as the precision in the system's color $\epsilon_{r-iEB}$ decreases for
OS$_{EB}$+AS3. 
{\sl Right:} Change in $S$ as $\epsilon_{RA}$ decreases in precision
 for OS$_{EB}$ and OS$_{EB}$+AS1.  
 \label{fig:signif_obs}}
\end{figure*}

Figure \ref{fig:signif_obs} left panel shows $S({T_{\rm eff\it A}})$
and 
$S({r-i})_{EB}$ as $\epsilon_{r-iEB}$ changes for OS$_{EB}$+AS3.
As $\epsilon_{r-iEB}$ decreases in precision, 
$S({r-i})_{EB}$ decreases from 1.0 to 0.6, 
while an increase in  $S({T_{\rm eff\it A}})$ from 0.6 to 0.8 is also seen, perhaps
{implying that $(r-i)_{EB}$ is a temperature indicator.}  
This is similar to the single star case but in the binary system
 \teff$_A$ remains relatively more important because the 
blended $(r-i)_{EB}$ provides less information
(see Fig.~\ref{fig:signif_obs_s} left diagram).

The right half of Figure \ref{fig:signif_obs} shows $S({R_A})$  as its
precision is decreased for OS$_{EB}$ and OS$_{EB}$+AS1.
{In the latter case, including an identified mode causes
$S(R_A)$ to rapidly decrease in value, 
as it becomes more poorly  determined.
For some observables ($T_{\rm BA}$, $M_{A,B} \sin^3 i$) $S$ increases slightly during this change.
At more than about 2\% error, $R_A$ has little role to play.
On the other hand,  for OS$_{EB}$, $R_A$ remains relatively important at all considered errors, simply because
there is no other observable in OS$_{EB}$ that provides the same type of information.}

\section{Determination of the fundamental parameters of the pulsating star}
\label{sec:dscuti_sgldbl}

The observables 
that showed most sensitivity to a change in their precisions for different sets for either
a single star or a binary system are $R$, \teff, the photometric colors and an identified mode.
In this section we study the uncertainties (Eq.~\ref{eqn:uncertainties}) 
in the global parameters of the (primary) pulsating
star as these observables are improved/included for both a single star and
a binary system.

\subsection{Parameter uncertainties}\label{sec:4.1}

\begin{figure*}
\includegraphics[width=0.5\textwidth]{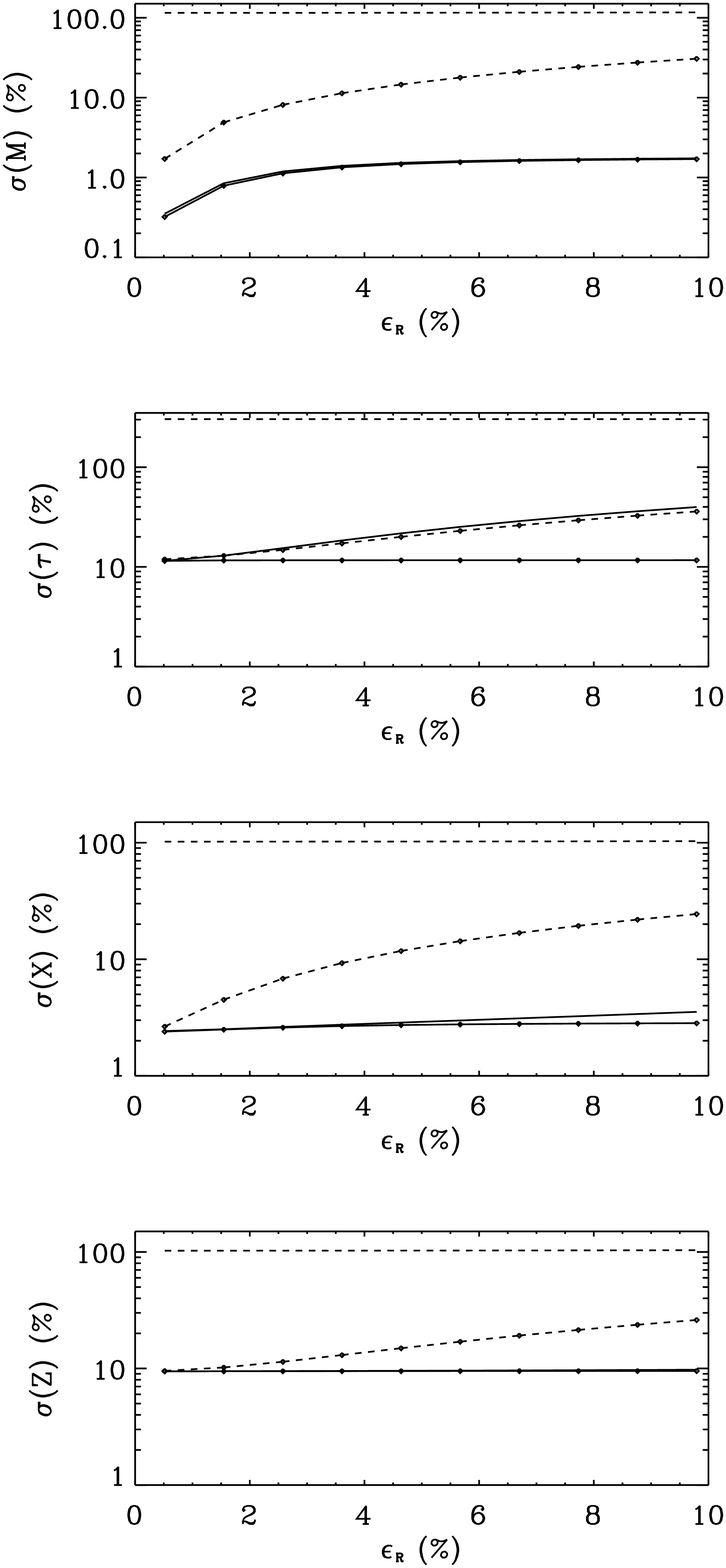}
\includegraphics[width=0.5\textwidth]{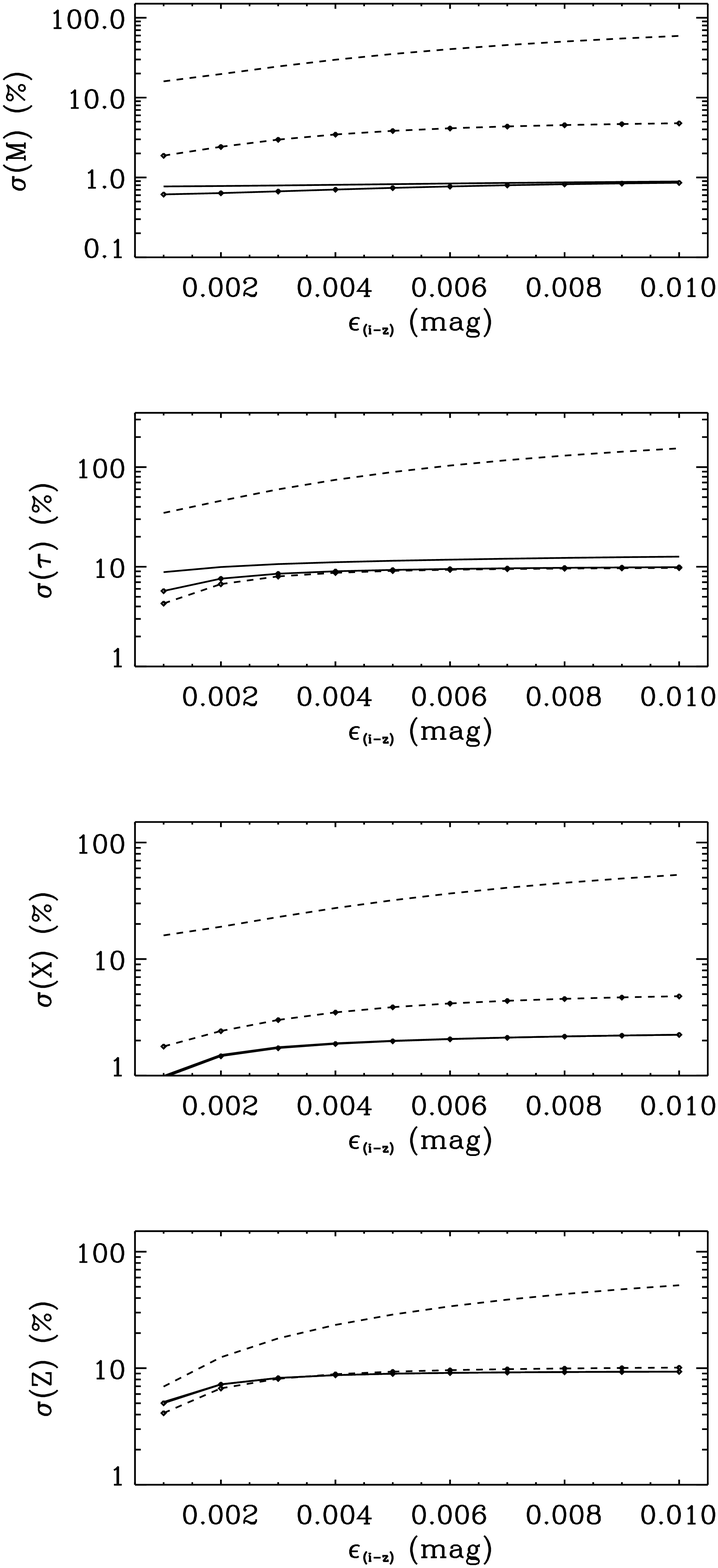}
\caption{Theoretical uncertainties in the stellar parameters 
as the precision in $\epsilon_{R}$ (left) and $\epsilon_{i-z}$ (right) increases,
for a single star (dashed lines) and a binary 
system (solid lines). 
The lines with dots represent the sets of observables
with seismic information.
The left panel shows the results without photometric constraints.
The y-axis for each parameter has the same scale
for comparing panels.
\label{fig:binsglpar}}
\end{figure*}

To successfully use a single identified mode as a ``seismic probe'', 
we need to know the fundamental stellar
parameters $M$, $\tau$, $X$, $Z$ and $v$ of the pulsating star with good precision.
Here we discuss only the uncertainties in the first four, because 
the determination of $v$ is independent of the 
determination of the others (the covariance off-diagonal elements $\sim0$) 
and independent of
most of the observable errors, except $i$ and $v_A\sin i$.
{All of the other parameters of the system ($M_B \sin^3 i$, $e$, etc.) 
are included for deriving the 
uncertainties, due to correlations that must be accounted for.
However, we do not discuss these uncertainties because we are only
interested in describing the pulsating star.
As we are dealing with the primary (A) pulsating star only, we drop the
subscript 'A'  on all of the quantities.}

Figure \ref{fig:binsglpar} shows the theoretical uncertainties in
$M$ (top), \age\ (second), $X$ (third) and $Z$ (lower panels) 
as we increase the precision in $R$ 
(left panels) and both of the   
 photometric colors (right panels). 
 The latter is denoted by $\epsilon_{i-z}$  and for the binary it implies
 the blended system's colors.
The dashed and solid lines represent the single and binary system respectively,
and the observable set with the identified mode has dots overplotted on the lines.
The results for OS (both OS$_S$ and OS$_{EB}$) 
and OS+AS1 are represented on the left panels, and 
those for  OS+AS2 and OS+AS3 (including photometric information) on the right.
Both the left and right panels show the same y-axes scales for comparison.

It is very clear that the oscillation mode plays a very different role in the single star and the 
binary system.
Excluding it from the single star constraints, without photometry, yields 
uncertainties $>$ 100\% and 
independent of $\epsilon_{R}$;
the constraints are inadequate to determine the stellar model.
Including the color 
constraints also yields large uncertainties ($>$ 30\%)
although not entirely independent of $\epsilon_{i-z}$.
When the mode is included, the uncertainties improve rapidly as the errors in 
both types of observables improve. 
In fact, $\sigma(M)$ becomes strictly dependent on $\epsilon_R$ as the error in the radius
improves to $<2$ -- 3\%.

There is little difference seen in the parameter uncertainties for the binary system when 
the oscillation mode is included/excluded as a constraint.
The  
binary system without any seismic data provides similar or better constraints on 
all of the parameters than the single star {\it with} an identified mode, except in some exceptional
cases (see below).
The fact that including a mode with the binary system constraints 
does not lead 
to improvements in the pulsating star's
parameter uncertainties indicates that the binary constraints alone may
provide sufficient constraints on the model 
to {\it use} the identified mode in a different way.

{For the binary system, we also tested the effect of improving the precision in \masini\ and \mbsini.
Setting both of these values to 0.01, 0.05, and 0.10 \msol\ (0.5\%, 2\% and 5\%) has little effect:
for $\epsilon_R$ = 2\%, $\sigma(M) = 0.3, 1.0$, and 1.5\% respectively with and without the identified
mode, and
for $\epsilon_R >$ 4\%, $\sigma(M)$ levels out to 0.3, 1.5, and 2.5\%.}

We note that for both $\sigma(Z)$ and $\sigma(\tau)$ for OS+AS3,
 in some cases the single star provides slightly better constraints than the binary.
This happens only for very precisely measured (unblended) colors, 
while including an oscillation mode as a constraint.
Finally, one should also note that $\sigma(X)$ $\gtrsim$ 10\% is not a good constraint.  The absolute value
of the parameter is 0.700, and a 0.070 error on this value produces no meaningful constraint on $X$.

\subsection{Parameter correlations}

\begin{figure*}
\includegraphics[width=0.49\textwidth]{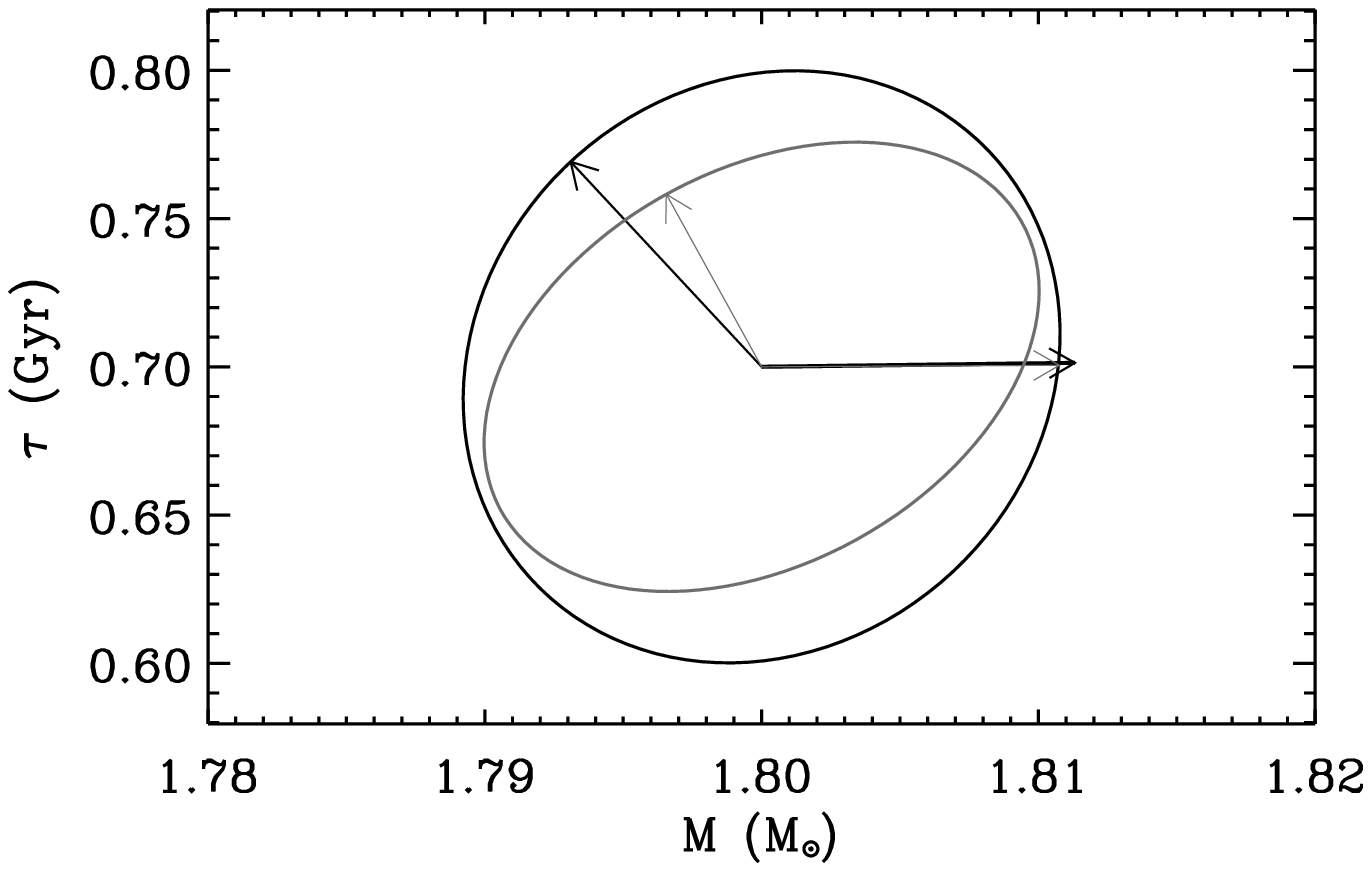}
\includegraphics[width=0.49\textwidth]{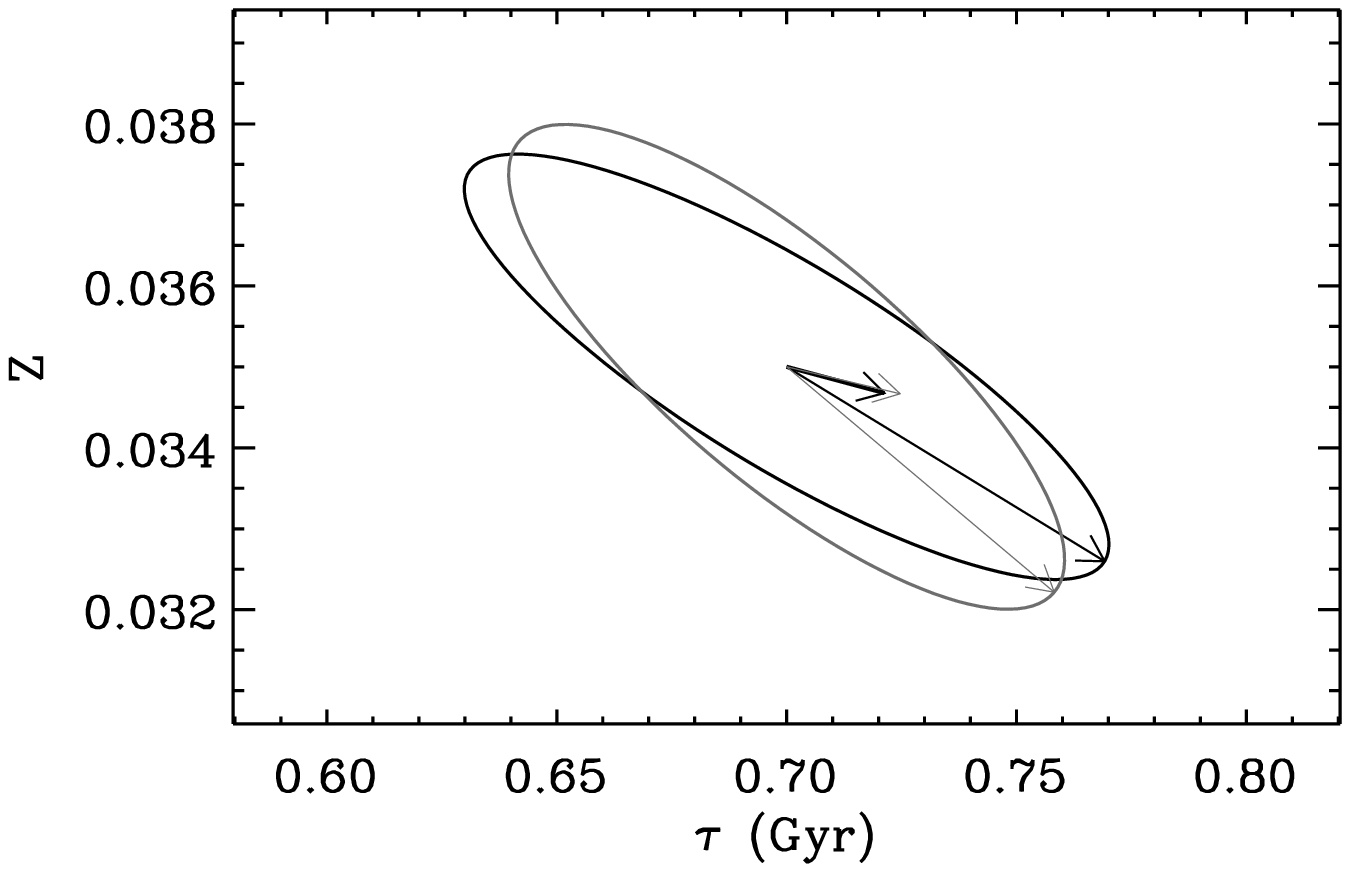}
\caption{Maximum projections of the two-dimensional error ellipses 
defining the region of parameter space where the solution lies for
the binary system, using OS$_{EB}$+AS2 (black) and OS$_{EB}$+AS3 (grey), and 
the observational errors given in the third column of Table~\ref{tab:obsers}.\label{fig:ellipses}}
\end{figure*}

The {\bf V} matrix from the SV decomposition describes the correlations among
the parameters, while the singular values scale each of the vectors $V_j$ to 
produce the n-dimensional error ellipse axes.
{In Figure \ref{fig:ellipses} we show projected contours of the 1-$\sigma$ error
box for $\tau$ versus $M$, and $Z$ versus $\tau$
using the vectors that yield the largest ellipses.}
The black/grey ellipses correspond to the 
sets of errors in the third column of Table~\ref{tab:obsers} for OS$_{EB}$+AS2/AS3. 
Including the oscillation mode does not reduce the error ellipse, 
and this reinforces the possibility
of using this extra information to learn something else about the star.

\subsection{H-R error box\label{sect:lt}}

\begin{figure*}
      \begin{center}
	\includegraphics[width = 1.\textwidth]
			{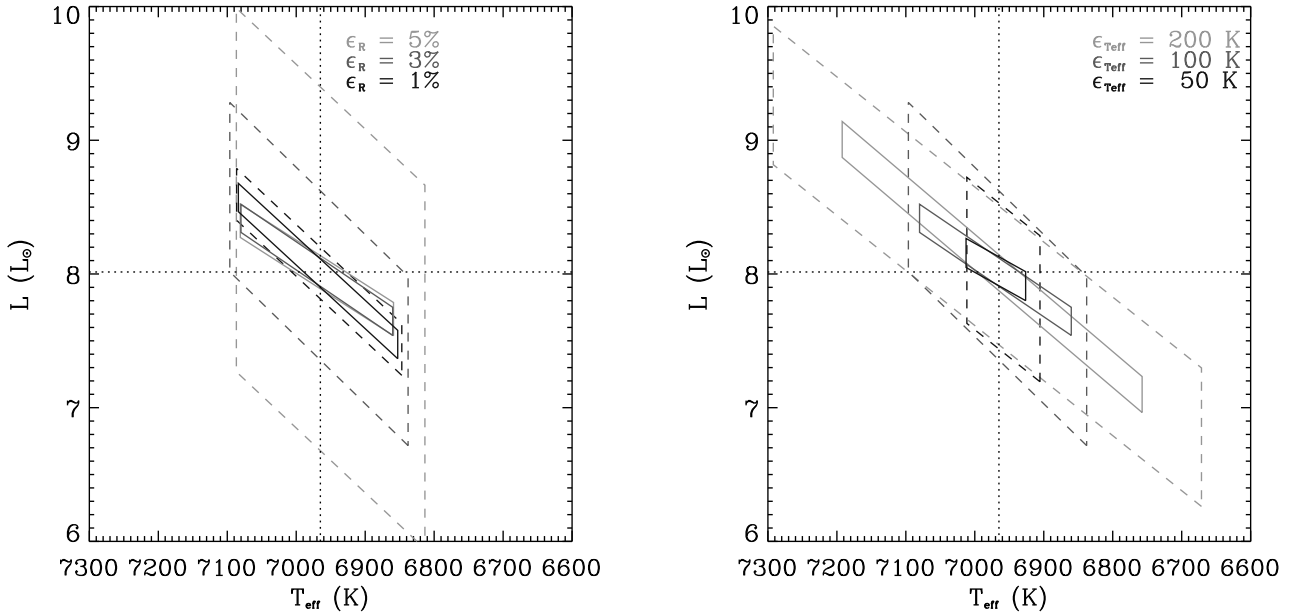}
      \end{center}
      \caption{\label{fig2} The theoretical error boxes for the model
      parameters of luminosity and 
	effective temperature.  The dashed/continuous lines represent the results for the 
	single star (OS$_S$+AS1)/binary system (OS$_{EB}$).
	The left/right panels show the results while reducing the error in observed
	$R$/\teff.}
    \end{figure*}

    Using equation~(\ref{eqn:dscuti_er}) we 
    calculate the theoretical uncertainties
    in the model effective temperature and model luminosity.
    Figure \ref{fig2} shows the approximate theoretical error boxes for these quantities
    for a single star (OS$_S$+AS1, {dashed lines})
    and a binary system (OS$_{EB}$ {solid lines}).  
    For the single star an identified mode is 
    included  
    because Fig.~\ref{fig:binsglpar} and Sect.~\ref{sec:4.1}
    showed that the parameters are
    not constrained if the identified mode 
    is not included,
    while 
    for the binary system no seismic data are included. 

    The error box shrinks significantly in the luminosity axis
    while reducing the error 
    in the radius for the single star ({left panel}).
     $\sigma(L_{\star})$ also changes slightly with
    $\epsilon_{T_{\rm eff}}$ (right panel), but its value is determined primarily by 
    the error in the radius (2\%). 
    The left panel shows that 
    interpolating between 1\% and 3\% for $\epsilon_{R}$
    produces a $\sigma(L_{\star})$ = 0.5 
    L$_{\odot}$ for $\epsilon_{R}$ = 2\%, as 
	shown in the right panel.

    As expected, {the error box shrinks in the model effective temperature axis due to 
    the reduced errors
     in the observational $T_{\rm eff}$ ({right panel}).}
    The observational $\epsilon_{T_{\rm eff}}$ of 200, 100, and 50 K produce model uncertainties   
    $\sigma (T_{\rm eff})$ of 250, 110, and 50 K.
    {We do not expect these values to be reproduced exactly because the observed
    \teff\ has a measurement error associated with it, and for large values of the
    error (\si 200K), other observables can dominate the determination of the derived model parameter
    of \teff.}
    However, the fact that these uncertainties are approximately reproduced provides
    evidence that the SVD method is valid.

    No identified mode is
    included with the binary system constraints (solid lines).
    The error box for the binary system does not shrink while 
    reducing the errors
    in the radius (left panel). 
    However, it shrinks when the error in the observed effective temperature 
    is reduced, reproducing accurately the observational  $\epsilon_{T_{\rm eff}}$
    of $\sigma(T_{\rm eff})$ = 200, 100, and 50 K.
    $\sigma (L_{\star})$ does not decrease in either panel,
    because the mass is well-determined for the binary system 
    and provides this strong constraint on $L_{\star}$.
    
    In all cases 
    the constraints provided by the binary system without an identified 
    mode
    on the parameters of the pulsating star
     are more 
    effective than those from the single star when an identified mode is 
    included.

\section{Incorrect mode identification}\label{sec:five}
Asteroseismology is only useful when the observed frequencies have been
correctly identified with their mode geometry.
Although techniques exist to do this, is there a way to test if this identification
is correct?
In this section we show that an incorrectly identified mode can be diagnosed if 
the pulsating star is in the binary system, but not as a single star.
We subsequently show that we can in fact constrain and/or identify the mode
correctly by studying the distribution of model solutions.

We have shown that the oscillation mode is an important observation 
to have (Sect.~\ref{sec:singlestar})
but including it as a constraint for the binary system did not
improve the pulsating star's parameter uncertainties 
nor shrink the error box in the H-R diagram (Sect.~\ref{sec:dscuti_sgldbl}).  
This implied that the mode is somewhat {redundant} information
for defining the global quantities, and
we concluded that
the observed oscillation frequency 
could possibly be used in a different way.
The oscillation mode occupies the top 
rows of the {\bf U$^{\rm T}$} matrix in the 
single star (Fig.~\ref{fig:dscuti_svd_sgla}) and in the binary case, indicating that it is an 
observable that tightly constrains the stellar model.
If the mode were incorrectly labeled, it should have a significant impact
on the stellar parameter solution. 
We have also shown that the observational constraints from the binary system
yield very well-determined parameters without using the 
oscillation mode (Figs.~\ref{fig:binsglpar}, \ref{fig:ellipses}). 
Therefore, we can compare the model solutions when 
we use a set of observables with and without the oscillation mode as a constraint.

\begin{figure*}
\includegraphics[width = 0.5\textwidth]{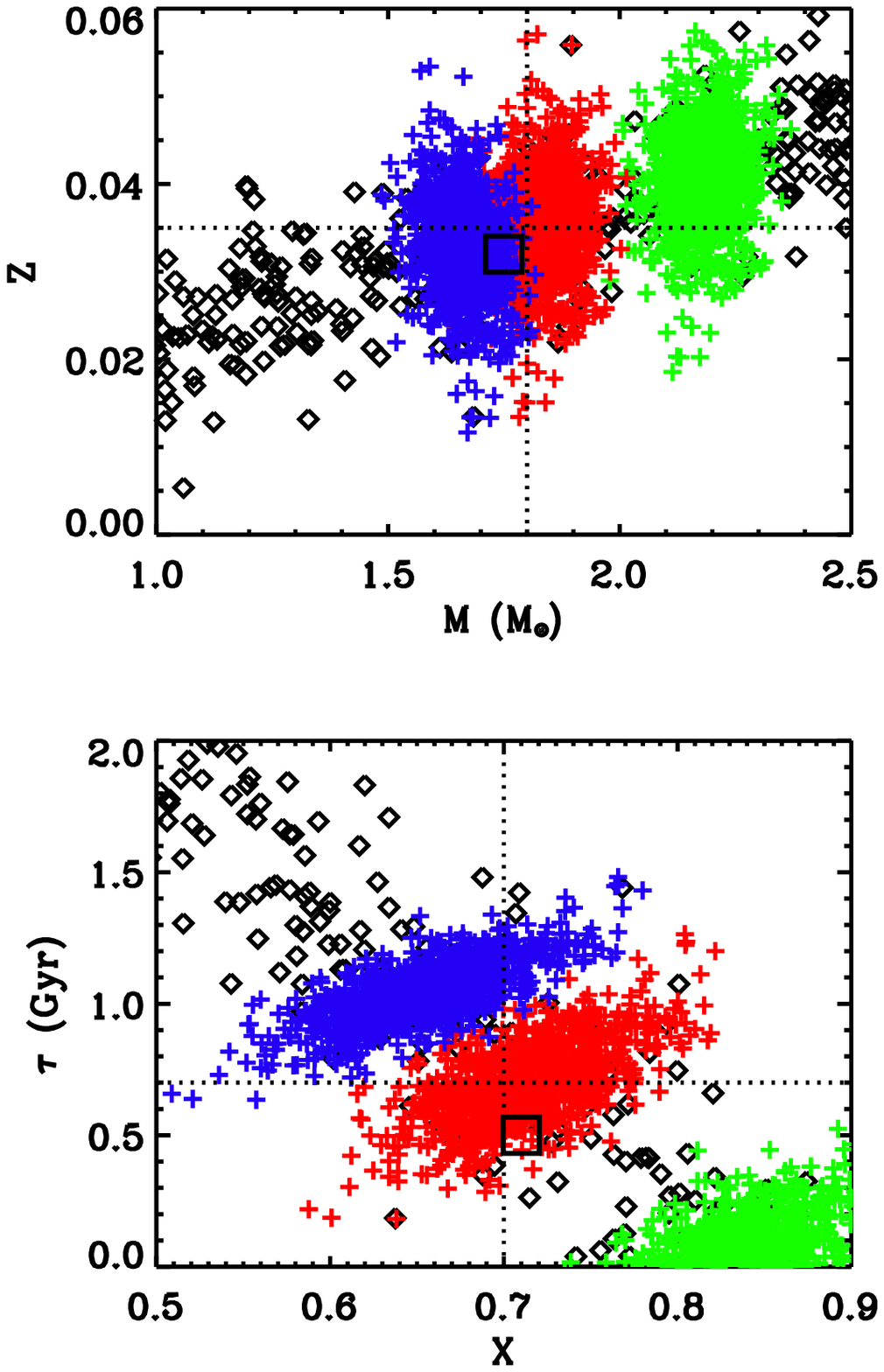}
\includegraphics[width = 0.5\textwidth]{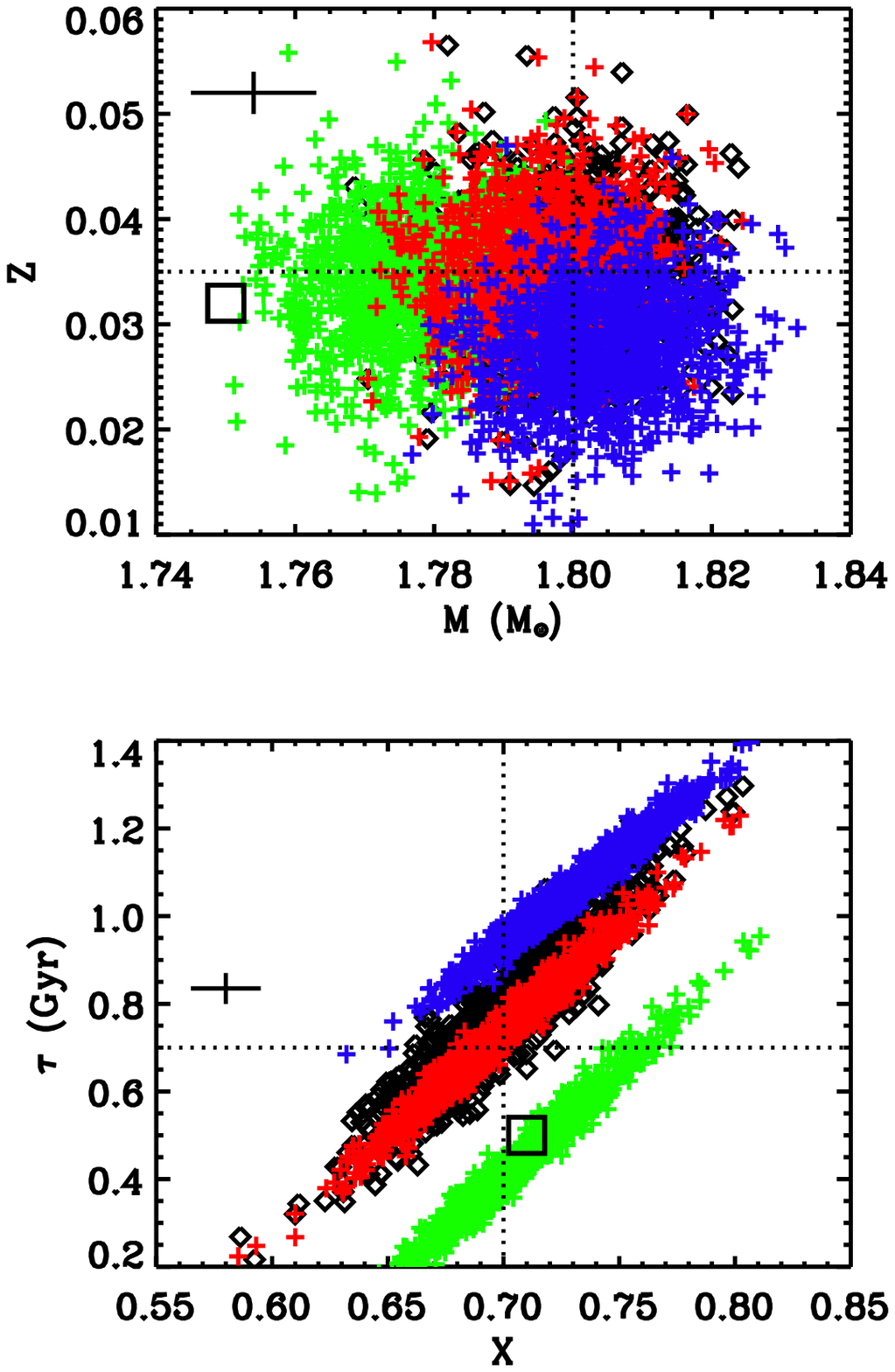}
\caption{Fitted parameters from simulations using OS$_{EB}$ and OS$_{EB}$+AS1.
The left/right panels show the results for the single star/binary.
The color-code is as follows: black --- no mode included; red, green 
and blue --- including a mode, with various identifications:
red --- correctly identified; blue --- incorrectly labeled degree; 
green  --- incorrectly labeled radial order.
The intersection of the dotted lines are the true values of the input parameters,  
the black square shows the values of the initial guess, and the
error bars on the right panel are the 1-$\sigma$ uncertainties for the binary 
system (see Fig.~\ref{fig:binsglpar}).
\label{fig:inversions}}
\end{figure*}

\begin{figure*}
\center{\includegraphics[width = 0.9\textwidth]{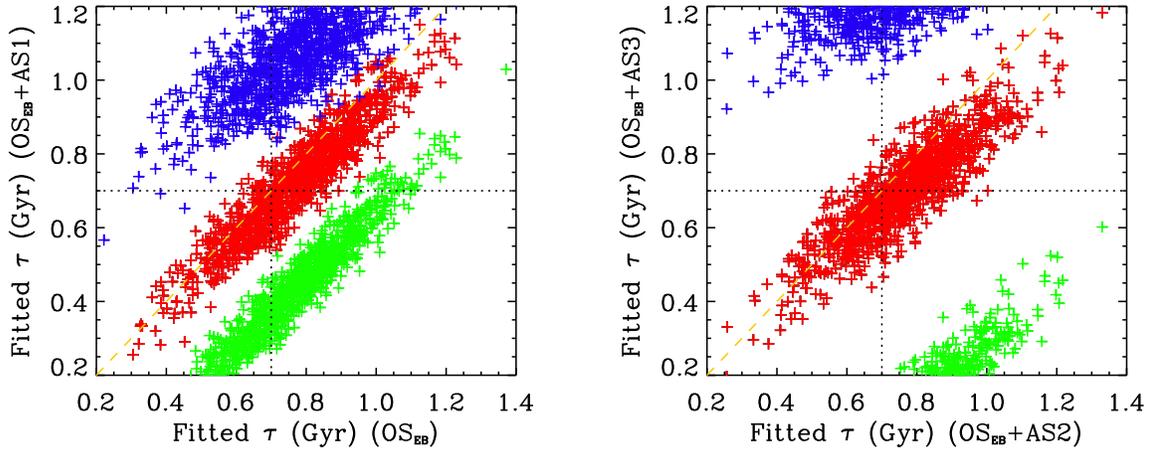}}
\caption{Fitted age from the simulations when a mode is (y-axis) and is not (x-axis) included as
an observational constraint.  The color code is the same as Fig.~\ref{fig:inversions}, and 
the left/right panels show the results for $OS_{EB}$+AS1/AS3.
\label{fig:fitage}}
\end{figure*}

We perform simulations to test the effect of 
recovering the stellar parameters when we identify a mode incorrectly.
The model that we test has the true input parameters ({\bf P}) 
that are given in 
Table~\ref{tab:parameters} with the corresponding model observables ({\bf B}) from
 Table~\ref{tab:obsers}.
We generate a set of ``real observations'' ($y_i$) by adding random errors to 
the true model observables: $y_i=B_i+\epsilon_ir_i$, where $r_i$ is a 
random number drawn from a Gaussian distribution with a mean of 0 and 
standard deviation of 1. 
To recover {\bf P} from $y_i$, we first make a guess of the 
solution {\bf P}$_0$ and then use equation~(\ref{eqn:dscuti_er}) to recover the true
parameters.  
We do this for four different scenarios for both the single star and 
the binary system, without including photometric information (OS$_{S/EB}$ and OS$_{S/EB}$+{AS1}):
 (1) using the non-seismic measurements {without} a mode (black diamonds
in Figure \ref{fig:inversions}, OS$_{S/EB}$),
 (2) using the non-seismic measurements {and} a correctly identified mode
(red crosses in Figures  \ref{fig:inversions} and \ref{fig:fitage}, OS$_{S/EB}$+AS1),
 (3) using the non-seismic measurements and an {\it incorrectly}
identified mode --- the wrong degree $\ell$ (blue crosses), and 
  (4) using the non-seismic measurements and an {\it incorrectly}
identified mode --- the wrong radial order $n$ (green crosses).

The recovered parameters for 10,000 realizations are shown in Figures \ref{fig:inversions} and \ref{fig:fitage}.
We show only the four parameters of the pulsating star discussed in Section~\ref{sec:4.1}.
The left/right panels show the results for the single star (OS$_{S}$/OS$_{S}$+AS1)/binary system (OS$_{EB}$/OS$_{EB}$+AS1).
In each of the panels the black square shows the initial guess of the 
corresponding parameters, while the dotted lines show the model values from Table~\ref{tab:parameters}.

For the single star the left panels of Figure~\ref{fig:inversions} show that 
the non-seismic observables (OS$_S$, black diamonds) 
do not constrain the parameters of the system.  
Once the mode is included (OS$_S$+AS1), the solution becomes tightly constrained to a small
range of parameter values.
When the mode is correctly identified (red crosses) the recovered 
parameter range is 
correctly centered on the intersection of the lines.
When the mode is incorrectly identified (blue or green crosses) the recovered
parameter range is incorrect. 
It is possible to discard the green solution (the incorrect radial order)
because the recovered values of $X$ are outside of the typically accepted values. 
However, it is not possible to determine 
if the blue or the red values
represent the true input parameters.
This means that we are either heavily reliant on obtaining a correctly-labeled mode,
or that the formal uncertainties are unrealistic as uncertainties because the systematic error is
completely ignored.

For the binary system the right panels of Figure~\ref{fig:inversions} show that 
the parameter uncertainties agree with 
those shown in Figure~\ref{fig:binsglpar}; 
the 1-$\sigma$ uncertainty is shown by the error-bar. 
The three solutions with the mode all lead to well-constrained parameter
ranges. 
The solution without the mode (black diamonds) 
can not be seen very clearly because the values coincide with the red solution, 
implying that it is the correct identification. 
The blue and the green solutions can be discarded simply because they 
are not in agreement with the black solution i.e. there is something incorrect
about the mode-identification.

To highlight the inconsistency in the solutions, Figure~\ref{fig:fitage} shows the
fitted age for the three solutions including the mode (y-axis) versus the solution
without the mode (x-axis).  
If the mode were correctly identified we would expect 
the solutions to approximately follow the bisector.
This is precisely what we see in both panels.  
When identifying the mode with  an incorrect degree (blue) or the incorrect radial order 
(green) there is a systematic offset, which implies an incorrectly labeled mode.
The difference between the left and the right panels is that we included 
photometric information in the latter (OS$_{EB}$+AS2/AS3): here the offset is much clearer
making it easier to detect the incorrectly-identified mode.
 

\begin{figure*}
\includegraphics[width = 0.5\textwidth]{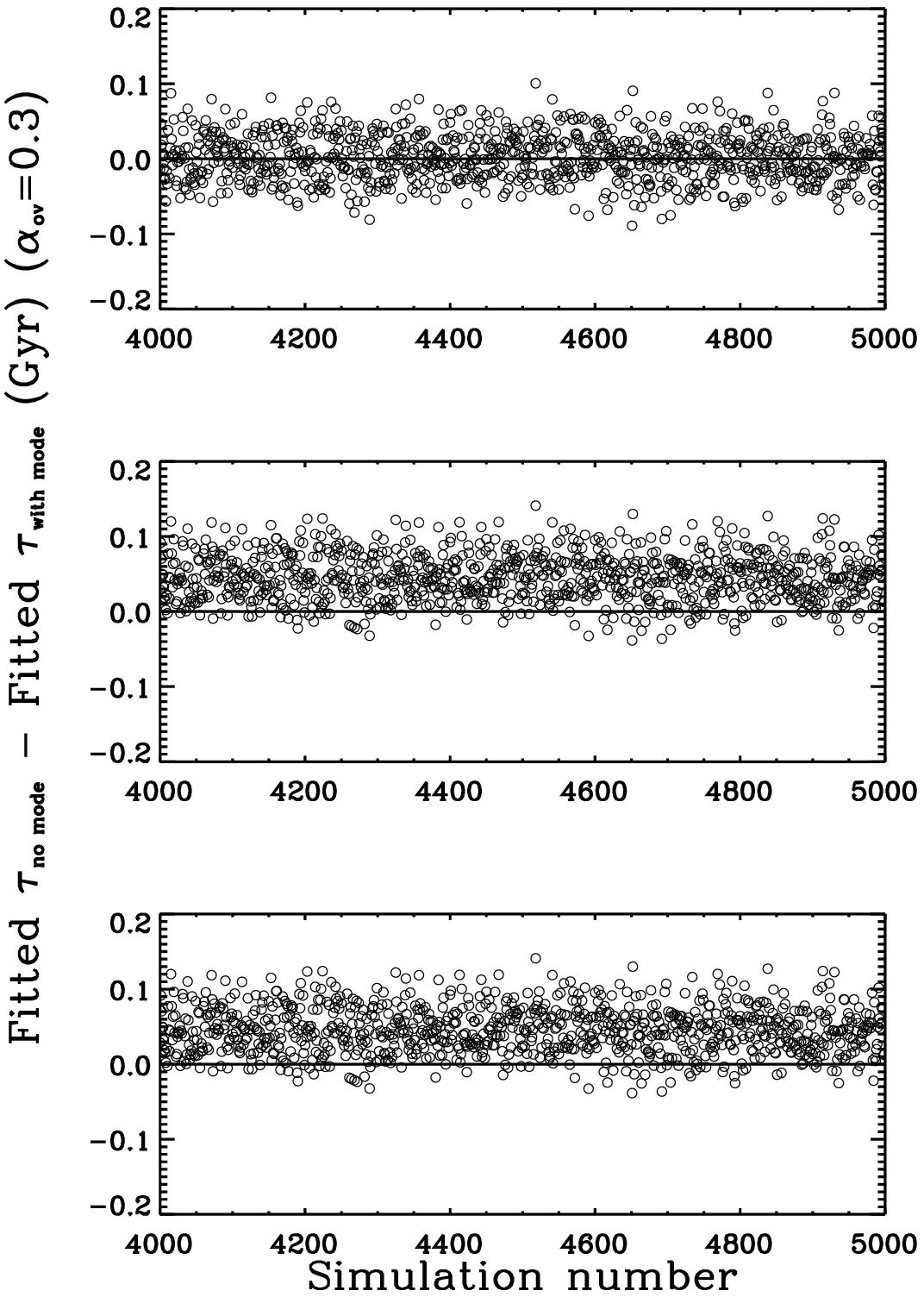}
\includegraphics[width = 0.5\textwidth]{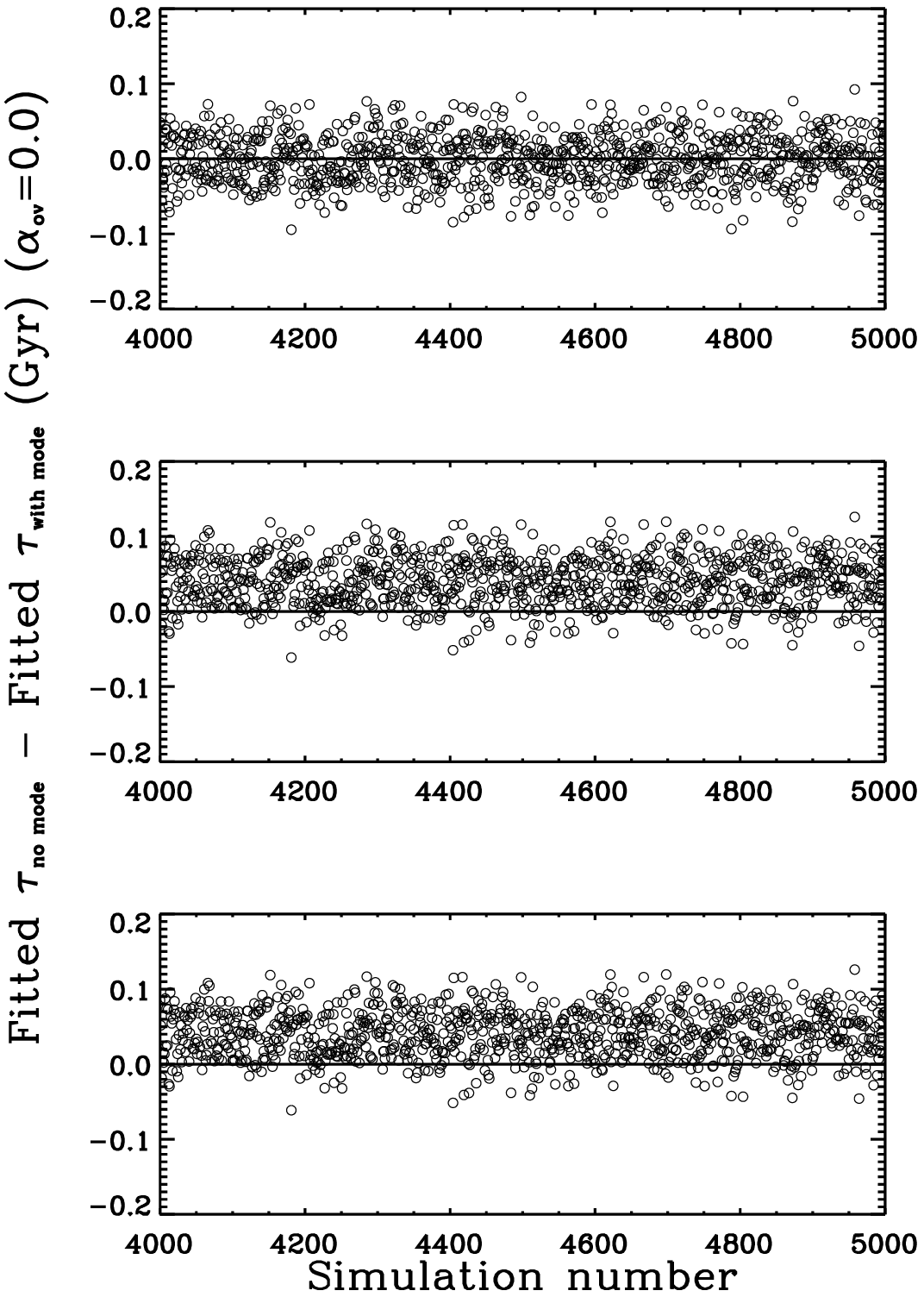}
\caption{Fitted age from simulations using OS+AS2 and OS+AS3.
The left/right panels show the results for inversions using the 
correct/incorrect model ($\alpha_{\rm ov} = 0.3/0.0$).
From top to bottom, the simulations assume a mode identification
of $(\ell,m,n)=(1,-1,9)$ (correct), $(0,0,9)$ (incorrect) and $(1,0,9)$ (incorrect).
\label{fig:inversions2}}
\end{figure*}

This question can be alternatively posed to ask if we can 
use the distribution of model solutions to actually identify
the oscillation mode.  
Figure~\ref{fig:inversions2} shows a representative 
sample of the distribution of model solutions of similar
simulations for the recovered age parameter.  
Each of the panels shows the recovered age without including
the mode~{\it minus}~the recovered age using an arbitrarily-identified mode. 
We show just the results for the most ambiguous solutions, because
all of the 
other mode identifications led to very clear discrepancies between the 
two solutions.  
For the simulations our observations come from a stellar model with 
a value of convective core overshoot parameter $\alpha_{\rm ov} = 0.3$.
We show results when we assume that the inverting model has 
a value of $\alpha_{\rm ov} = 0.3$ (correct) and 0.0 (incorrect).
The top panel shows the results when the mode identification is correct
$(\ell,m,n)=(1,-1,9)$.
The middle and lower panels show the results when the mode is labeled incorrectly:
$(0,0,9)$ and $(1,0,9)$.
In both panels we can see that even with some errors in our assumptions 
about the model, we can safely constrain, or even label, 
the mode with the correct identification from the top panel. 
This does not imply that the identiÞed mode is insensitive to $\alpha_{\rm ov}$, simply that by comparing solutions from the same model any possible systematic error is eliminated.


\section{Conclusions}\label{sec:conclusions}
{Seismology of $\delta$ Scuti stars has had limited success due to the inability to obtain a unique model of the star and more importantly the difficult task of mode-identification due to rotation, evolution and 
a deficiency in our understanding of mode-excitation mechanisms.
A possible method for overcoming these obstacles, as suggested by many authors, 
is to study pulsating stars in detached eclipsing spectroscopic binary systems.
In this paper we used singular value decomposition (SVD) to quantify the
advantages of studying pulsating stars in eclipsing binaries over single field stars,
by comparing the parameter uncertainties and the ability to detect an incorrectly-labeled 
mode in both cases.}

We have demonstrated that SVD is a powerful tool to assess the 
parameter constraints in various systems 
considering the available measurements. 
This method assumes that we are relatively close to the 
parameter solution so that the model derivatives can be described locally as linear. 
This implies that the results are not extendable to other
values where we can no longer extrapolate linearly to 
estimate observables.  
We considered the set of parameters shown in 
Table~\ref{tab:parameters} and deviating much from these values will 
require new calculations. 
This work also imposed various assumptions and they  must be taken into account.
For example, the derivatives are calculated from the ASTEC \citep{jcd08} stellar code,
and they may vary (hopefully only slightly) from one code to the next.

We have shown that as the measurement errors on the photometric information are varied, the roles of 
the observables change especially for the single star.  In particular,
$R$ and $(r-i)$ play similar roles for determining $M$,  and 
$(i-z)$ and [M/H] for determining $Z$.
The distance $d$ is uniquely determined by $\pi$ when no photometric information
is included, while the colors and magnitude contribute to  
determining $d$ when these are included, primarily due to the error on the parallax.  
When no oscillation mode is included, then all of the observables are  
important for constraining the parameters even  poorly, and
the inclusion of a correctly identified mode replaces the role of $R$ for determining $M$.

For the binary system  the photometric information does not have as much
impact as it does for the single star, partially because these observations 
are blended values resulting from the two components.
However on reaching precisions of $\sim 0.003$ mag, then the 
photometric observables begin to dominate the roles of
$R_A$, \teff$_{A}$, [M/H], and \trat\ for 
determining the stellar parameters.
In particular,  $(r-i)_{EB}$ yields similar information to \teff$_A$. 

For a single star system correctly identifying  
an oscillation mode is necessary to reduce the parameter uncertainties
to values that are useful.  
On the other hand, a binary system contains enough information, without having
to use the identified mode to constrain the stellar parameters well. 
In fact, including the identified mode does not improve the parameter uncertainties,
emphasizing its utility for learning something else about the star.
This same trend can be seen for the error box in the H-R diagram: 
it does not shrink for the binary system when 
the mode is included or the errors in $R_A$ are improved.
For the single star, however, improving $R$ leads directly to shrinking the 
error box in the luminosity axis only when a correctly 
labeled mode is included.

Other parameter values were also considered in this study; we 
varied the primary mass of the star (1.80 \msol\ and 2.50 \msol), 
the mass ratio of primary to secondary (1.10 and 1.60), 
the evolutionary stage of the primary star (central hydrogen mass fraction 
of 0.50 [MS] and 0.15 [end of MS]), and 
the metallicity of both stars (0.020 and 0.035). 
For a single star identifying a mode  has more impact on the reference model than
the more evolved model and 
for both the single star and the binary system the higher metallicity model benefited more from the 
photometric information.
However, changing the size of the primary mass 
did not change the uncertainty dependencies  \citep{cre08}.

Because mode-identification in $\delta$ Scuti (and other) stars is 
a complicated task, we addressed whether
an incorrectly-labeled mode could be correctly diagnosed using
the observational constraints provided by the single star and the detached eclipsing
binary system.  
We used simulations to test the effects of recovering 
stellar parameters with arbitrarily-identified oscillation modes.
We found that for the single star, it is necessary to identify the mode unambiguously, 
because the recovered solutions using an incorrect or a correct 
mode identification are not distinguishable.
Meanwhile for the binary system, 
an incorrectly-identified mode can
be diagnosed by simply comparing the parameter solutions when
the mode is included and not. 
If the mode is correctly identified, the 
solutions are in agreement. 

Taking this discussion slightly further, in the binary system we also showed that we could in fact
tightly constrain (and even identify correctly) 
the pulsation mode parameters  
by comparing the distribution of model
solutions with and without a mode. 
The model solutions are in agreement when the mode is correctly 
identified even if the assumptions on the interior physics are slightly incorrect. 
This type of test and  this method of mode-identification 
has not yet been done with available observational data.

It is clear that studying pulsations in binary systems has its complexities
observationally, however, the extra effort is
definitely worth it. 
Apart from the advantage of using the screening effect of the pulsations during eclipse
to help identify the mode, 
this study emphasizes
the power that the binary measurements have for determining
fundamental parameters and subsequently constraining and even 
identifiying the mode.

In this paper we use the SVD technique to compare the constraints for single 
stars and binary systems.
We note, however, that with the flood of data from space-based missions, 
this type of study can be subsequently extended to investigate 
which detached eclipsing binary systems 
provide the best astrophysical laboratories.





\acknowledgments

Part of this work was supported by a graduate student {\it Newkirk Fellowship} at 
the High Altitude Observatory/National Center for Atmospheric Research, Boulder, CO, USA.
This work was also supported by the European Helio- and Asteroseismology
      Network (HELAS), a major international collaboration funded by the 
      European Commission's Sixth Framework Programme.
      We thank the referees for comments and suggestions 
      that greatly improved the presentation and the scientific
      content of the manuscript.

\end{document}